\documentclass[pdflatex,sn-mathphys-num]{sn-jnl}

\usepackage{amsmath,amssymb,amsfonts}
\usepackage{graphicx}
\usepackage{lineno}
\usepackage{longtable}
\usepackage{array}
\usepackage{ragged2e}
\usepackage{hyperref}
\usepackage{enumitem}

\title{Generalized Normal Constraints (GNC): A Complete Geometric Generalization of the NNC Method}

\author*[1]{\fnm{Achille} \sur{Messac}}
\email{messac@howard.edu}

\author[1]{\fnm{Blayne} \sur{Montaque}}
\email{blayne.montaque@bison.howard.edu}

\affil[1]{\orgdiv{Department of Mechanical Engineering},
\orgname{Howard University},
\orgaddress{\city{Washington}, \state{DC}, \country{USA}}}

\abstract{
\unboldmath
This paper presents a unified geometric, mathematical, and computational framework for the generation of the \textit{complete} admissible Pareto frontier. Several existing methods are structurally unable to capture the complete admissible Pareto frontier. These include widely used methods such as the weighted sum, the Normal Boundary Intersection (NBI) method, and the Normalized Normal Constraint (NNC) method. NNC and NBI, which share the same Pareto-generation grid construction, are structurally unable to capture 50\% of the admissible Pareto region for tri-objective problems. More generally, for an $n$-objective problem, the admissible capture fraction decreases factorially as $1/(n-1)!$, and the corresponding missed fraction increases to $1-1/(n-1)!$. By contrast, the newly developed Generalized Normal Constraint (GNC) method introduced in the present work is structurally capable of capturing the complete admissible Pareto frontier.

The proposed GNC method is formulated for general $n$-objective optimization problems and is developed through a unified geometric, mathematical, and computational framework supported by computational examples. Multiobjective optimization plays an important role in a broad range of applications, including economics, product design, and engineering management. Accordingly, the ability of a Pareto-generation method to generate a representative subset spanning the \textit{complete} admissible Pareto frontier is of fundamental importance for multiobjective optimization.
}

\keywords{Multiobjective optimization; Pareto frontier generation; Generalized Normal Constraints (GNC); Normalized Normal Constraints (NNC); Pareto frontier completeness; Objective-space geometry; hypercube projection}

\begin{document}

\maketitle


\section*{Nomenclature}

\begin{center}
\begin{tabular}{ll}
\hline
\textbf{Symbol} & \textbf{Description} \\
\hline

$n$ & Number of objectives \\

$n_d$ & Number of grid divisions used for Pareto generation \\

$x$ & Design-variable vector \\

$\mu(x)$ & Objective-function vector \\

$\mu_i(x)$ & $i^{th}$ objective function \\

$\bar{\mu}(x)$ & Normalized objective vector \\

$\bar{\mu}_i(x)$ & Normalized value of objective $i$ \\

$\mu^{U}$ & Utopia-point vector \\

$e_i$ & $i^{th}$ standard basis vector \\

$E$ & Vector of ones \\

$\alpha$ & Grid-generation coordinate vector \\

$\mathcal{C}_n$ & $n$-dimensional normalized objective-space hypercube \\

$A_i$ & Anchor point corresponding to objective $i$ \\

$h_i$ & Hexagon vertex associated with GNC construction \\

$C_i$ & Cube vertex that lies on axis $\bar{\mu}_i$ for GNC construction \\

\hline
\end{tabular}
\end{center}

 

\section{Introduction}

Multiobjective optimization problems occur in areas ranging from engineering design to economics. They involve competing objectives that must be considered simultaneously. Unlike single-objective optimization problems, which generally seek one optimal solution, multiobjective problems lead to a set of Pareto-optimal solutions representing different tradeoffs among the objectives.

The ultimate objective of any deterministic Pareto-generation method is to produce a representative approximation of the \textit{complete} admissible Pareto frontier. Incomplete capture may prevent decision makers from identifying important tradeoff solutions and may therefore lead to suboptimal design decisions. Consequently, the ability to generate the complete admissible Pareto frontier is of fundamental importance.

To attain the key objective of \textit{complete} admissible Pareto frontier capture, the present work departs from traditional thinking. Rather than beginning with the design of a new Pareto-generation algorithm, the proposed framework first identifies \textit{the invariant geometric structures induced by the multiobjective optimization problem itself}. This approach resulted in the natural emergence of the proposed Generalized Normal Constraint (GNC) methodology from these structures --- where the generation of the \textit{complete} admissible Pareto frontier is a central pursuit.

\noindent \textit{The development presented in this paper is guided by four foundational questions. Collectively, these questions define the scientific progression of this paper and provide the framework through which the proposed GNC methodology naturally emerges.}

\begin{enumerate}
    \item What invariant geometric structures underlie complete admissible Pareto frontier generation?
    \item\ How can these invariant structures be used to understand the generation of Pareto frontiers?
    \item What do these invariants reveal about the capabilities and limitations of existing methods?
    \item How can these insights be exploited to develop and validate a method capable of complete admissible Pareto frontier generation?
\end{enumerate}

The remainder of this Introduction briefly reviews the relevant literature in the context of these questions and identifies the research gap that motivates the proposed approach. The subsequent sections then address each question in sequence.

In attempting to uncover the geometric structures that underlie the pursuit of the \textit{complete} admissible Pareto frontier, we begin with the end in mind. This perspective naturally leads to the question: "Where are the sought-after Pareto solutions located?". This question reveals that these structures --- fully examined in the present work --- include:

\begin{itemize}
    \item \textit{Normalized Objective Space} 
    \item Utopia Hypercube
    \item Pareto Capture Polyhedron 
    \item Anchor Points
    \item Utopia Point
    \item Pseudo Nadir Point
    \item Utopia Hyperplane 
    \item Pareto Capture Polygon
    \item Pareto Capture Projection
    \item Pareto Capture Grid
    \item Pareto Capture Fibers
\end{itemize}

The geometric structures identified above are invariant properties of the multiobjective optimization problem itself. Consequently, they are independent of any particular Pareto-generation methodology and provide a common geometric framework for understanding existing methods as well as developing new ones. Section 2 develops these invariant structures and demonstrates how they naturally lead to the proposed GNC methodology. The grid and fiber constructions also exhibit connections with classical lattice and coordination-sequence concepts \citep{ConwayGuy1996BookOfNumbers,ConwaySloane1997Coordination}, although the present development derives the required structures directly from the multiobjective geometry.

The multiobjective optimization literature is vast, and the applications have been equally broad over the past half century. Advances in computing have redefined the realm of the possible. We focus here on the approach of Pareto frontier generation, which provides the means to make effective decisions.   

Notable methods include the Normal-Boundary Intersection (NBI) method \citep{DasDennis1998NBI}, the Normal Constraint (NC) method developed by \citep{MattsonMessac2004NC}, and the Normalized Normal Constraint (NNC) method \citep{MessacIsmailYahayaMattson2003NNC}.
Notable NNC modifications include \citep{MartinezSanchisBlasco2007ModifiedNNC,  SanchisMartinezBlascoSalcedo2008ENNC, 
HancockMattson2013SmartNC, 
MottaAfonsoLyra2012ModifiedNBINC, 
ErfaniUtyuzhnikov2010DSD,
ErfaniUtyuzhnikovKolo2013ModifiedDSD,
UtyuzhnikovFantiniGuenov2009ParetoSet}.

These approaches have proven highly effective; however, important common limitations remain, including: \textit{incomplete} frontier capture, uneven distribution in certain geometric regions, dependence on objective ordering, redundant or non-Pareto points, and computational scaling as the number of objectives increases. 

Other publications that modified NBI and NC include the application to power system optimization \citep{RahmaniAmjady2018ImprovedNNC,LinLiuLiLuYanLiu2017NNC}; and comparative studies involving numerical implementations of NNC \citep{MartinezGarciaNietoSanchisBlasco2006GAvsGN}. Additionally, others have compared NBI and enhanced NNC formulations in engineering applications \citep{LogistVanImpe2012NBIENNC},    and developed interactive NBI and enhanced NNC procedures have also been developed for progressive exploration of the criteria space in optimization and optimal-control problems \citep{VallerioVercammenVanImpeLogist2015InteractiveNBIENNC}.
 
Evolutionary and stochastic methods provide another class of approaches for the generation of Pareto frontiers. Herrero et al. proposed the ev-MOGA algorithm for generating smart-distributed Pareto fronts \citep{HerreroReynosoMezaMartinezBlascoSanchis2014EvMOGA}. Martinez-Iranzo et al. investigated stochastic solvers for applied Pareto multiobjective optimization \citep{MartinezIranzoHerreroSanchisBlascoGarciaNieto2009Stochastic}. A broader context for evolutionary and multiobjective optimization methods is provided by survey and tutorial papers \citep{MarlerArora2004Survey,EmmerichDeutz2018Tutorial,HuaLiuHaoJin2021IrregularParetoSurvey, MartinsNing2021EngineeringDesignOptimization}.

Mueller-Gritschneder et al. \citep{MuellerGritschnederGraebSchlichtmann2009BoundedPareto} address the important problem of achieving a good representation of bounded Pareto frontiers through an adaptive sampling strategy through successive approximation. Eichfelder \citep{Eichfelder2009AdaptiveScalarization} introduced an adaptive scalarization method based on the Pascoletti--Serafini formulation and sensitivity analysis. The method adaptively controls the scalarization parameters to produce nearly equidistant approximations of the efficient set and is formulated for nonlinear multiobjective problems with general cone-induced partial orderings. The present work differs from the preceding two by first identifying \textit{the invariant geometric structures induced by the multiobjective optimization problem itself}, from which the GNC methodology naturally emerges. It further derives a closed-form inverse-factorial expression quantifying the structural incompleteness of NNC and establishes a framework that is structurally capable of generating the \textit{complete} admissible Pareto frontier.

Adaptive weighting strategies have been proposed to improve point placement and distribution, including the adaptive weighted-sum method of \citep{KimDeWeck2005AWS}. Other scalarization approaches have sought to improve access to nonconvex efficient frontiers, including $\epsilon$-constraint-based methods \citep{ChircopZammitMangion2013EConstraint} and alternative scalarization formulations developed specifically for nonconvex multiobjective problems \citep{GhaneKanafiKhorram2015Scalarization}.

These works have continually improved the Pareto frontier generation endeavor by creating new methods or improving old ones. One of the longstanding challenges in multiobjective optimization has been the generation of the \textit{complete} admissible Pareto frontier. Much less attention has been devoted to identifying \textit{the invariant geometric structures induced by the multiobjective optimization problem itself}. The present work instead focuses on identifying and exploiting these invariant geometric structures to achieve \textit{complete} admissible Pareto capture.
 
 Building upon the invariant geometric structures developed in Section 2, the proposed GNC methodology is formulated as their natural computational realization. The resulting method is subsequently validated through representative benchmark problems that confirm the theoretically expected complete admissible Pareto representation, geometric consistency, and practical computational performance.

The remainder of the paper follows these four questions in sequence. Section 2 establishes the invariant geometric framework, Section 3 uses this framework to analyze the capabilities and limitations of existing methods, Section 4 develops the proposed GNC methodology, and Section 5 provides computational validation. Through this progression, the proposed GNC methodology emerges naturally from the underlying geometric development.

 
\section{Fundamental Geometric Invariants of Multiobjective Optimization}

This section introduces the fundamental geometric properties that are invariant with respect to the particular Pareto-generation method employed. These invariant definitions and properties arise directly from the multiobjective optimization problem itself and therefore hold independently of the Pareto-generation methodology under consideration. As shown in the following developments, they naturally lead to the formulation of the proposed Generalized Normal Constraint (GNC) method. In the following, we present a sequence of mutually supporting geometric constructions that arise naturally from the multiobjective optimization problem itself.\\

The geometric development of the paper proceeds through the following sequence of invariant constructions:

\begin{enumerate}[label=\arabic*.]
  \item Normalized Objective Space --- \textit{Provides the natural geometric framework.}
  \item Utopia Hypercube --- The unit hypercube in normalized objective space.
  \item Pareto Capture Polyhedron (PCP) -- \textit{Defines the complete admissible Pareto region.}
  \begin{itemize}
      \item Pareto Capture Polyhedron for Convex Frontiers (PCP-C)
  \end{itemize} 
  \item Utopia Hyperplane 
  \item Pareto Capture Polygon (PCPg)--- \textit{Defines the complete admissible projection domain.}
  \begin{itemize}
      \item NNC Capture Polygon 
      \item GNC Capture Polygon 
  \end{itemize} 
  \item Pareto Capture Projections --- Maps admissible points onto the projection domain.
  \item Pareto Capture Grid  --- \textit{Provides an intrinsically uniform search lattice.}
  \item Pareto Capture Fibers --- \textit{Reduces projection computation \text{\;from Order}
$O\!\left(n_d^{\,n}\right)$ \;\text{to Order}\;
$O\!\left(n_d^{\,n-1}\right)$.}
  \item Method Independence of the above Definitions and Properties
\end{enumerate}

It is worth emphasizing that these structures emerge naturally by following the inherent geometric properties of the multiobjective optimization problem itself. Collectively, these geometric constructions naturally lead to the formulation of the GNC methodology, whose inherent geometric structure is capable of complete admissible Pareto frontier capture.


\subsection{Multiobjective Problem Statement and Formulation}

This subsection establishes the mathematical formulation of the generic multiobjective optimization problem that underlies all subsequent geometric developments. We begin by expressing the generic multiobjective optimization problem in the following form.

$\quad \textit{Problem} \; PU_0$

\begin{linenomath}
\begin{equation}
\min_{x} \; \mu(x) , \quad \mu(x) =
\{
\mu_1(x) \;
\mu_2(x) \;
\ldots \;
\mu_n(x)
\}^T \quad 
(n \geq 2)
\end{equation}
\end{linenomath}
subject to
\begin{linenomath}
\begin{equation}
g_j(x) \leq 0  \quad (1 \leq  j \leq p)
\end{equation}
\end{linenomath}
\begin{linenomath}
\begin{equation}
h_k(x) = 0 \quad (1 \leq  k \leq q)
\end{equation}
\end{linenomath}
\begin{linenomath}
\begin{equation}
x_{li} \leq  x_i \leq x_{ui} \quad 
(1 \leq  i \leq n_{x})
\end{equation}
\end{linenomath}

For convenience, the quantities introduced in this subsection are summarized below. 
\smallskip

\begin{tabular}{ll}

$n$ & Number of objectives \\

$(\cdot)_i$ & $i$th component of the corresponding vector\\
 
$x$ & Design-variable vector $\left(x\in\mathbb{R}^{n_x}\right)$\\

$\mu(x)$ & Objective-function vector $\left(\mu(x)\in\mathbb{R}^{n}\right)$ \\

$g(x)$ & Behavioral inequality constraint vector $  (g(x)\in\mathbb{R}^p$) \\

$h(x)$ & Behavioral equality constraint vector $  (h(x)\in\mathbb{R}^q$) \\

$x_l$ and $x_u$ & Lower- and upper-bound vectors $\left(x_l,x_u\in\mathbb{R}^{n_x}\right)$\\
 
\end{tabular}


\subsection{Anchor and Nadir Points, and Normalized Objective Space}

We first define the $i$th \textit{Anchor Point}, $A_i$, and then introduce the \textit{Normalized Objective Space}, both of which play central roles in the developments that follow. Specifically, the $i$th anchor point is obtained as the solution of the following single-objective optimization problem.

$\quad \textit{Problem} \; PU_i$
\begin{equation}
\min_{x} \; \mu_i(x)  \quad (1 \leq  i \leq n)
\end{equation}

\noindent subject to

\begin{linenomath}
\begin{equation}
g_j(x) \leq 0  \quad (1 \leq  j \leq p)
\end{equation}
\end{linenomath}

\begin{linenomath}
\begin{equation}
h_k(x) = 0 \quad (1 \leq  k \leq q)
\end{equation}
\end{linenomath}

\begin{linenomath}
\begin{equation}
x_{li} \leq  x_i \leq x_{ui} \quad 
(1 \leq  i \leq n_{x})
\end{equation}
\end{linenomath}
Let $x^{i*}$ denote the solution to $\textit{Problem}\;PU_i$. The corresponding anchor point is then
\begin{linenomath}
\begin{equation}
A_{i} \equiv \mu^{i*}= \mu(x^{i*})  
\quad \quad (A_{i}\in\mathbb{R}^n, \; x^{i*}\in\mathbb{R}^{n_x})
\end{equation}
\end{linenomath}

We then express the utopia point in \textit{physical objective space} as
\begin{linenomath}
\begin{equation}
U_p = \{ 
A_{11} \;
A_{22} \;
\ldots \;
A_{nn}
\}^T  \quad (U_p\in\mathbb{R}^n)
\end{equation}
\end{linenomath}

\noindent where $A_{ik}$ denotes the $k$th component of the $i$th anchor-point vector. Similarly, the \textit{Pseudo-Nadir Point} vector, hereafter referred to as the \textit{Nadir Point}, is defined as

\begin{linenomath}
\begin{equation}
N_p = \{ 
N_1 \;
N_2 \;
\ldots \; 
N_n
\}^T,  \; \quad \text{with} \quad
N_i =  \max_{k=1 \; \ldots \; n} \; A_{ki} \;
\quad (N_p\in\mathbb{R}^n)
\end{equation}
\end{linenomath}
Using the utopia and nadir points defined above, the normalized objective vector is denoted by
\begin{linenomath}
\begin{equation}
\label{eq:NormObjCom}
\bar{\mu}
=
\left(
\bar{\mu}_1,
\bar{\mu}_2,
\ldots,
\bar{\mu}_n
\right)^T, \quad     
\left( 0 \le \bar{\mu}_i \le 1 \right) \;
\forall \; i
\end{equation}
\end{linenomath}
In \textit{normalized objective space}, the utopia point becomes
\begin{linenomath}
\begin{equation}
U_p 
\equiv 
\bar{\mu}_p 
=
\mathbf{0}
\end{equation}
\end{linenomath}
In \textit{normalized objective space}, the $k$th anchor point has all components equal to one except for its $k$th component, which is zero. Thus,
\begin{linenomath}
\begin{equation}
A_k
=
\left(1,\ldots,1,0,1,\ldots,1\right)^T 
\end{equation}
\end{linenomath}
%


\subsection{Pareto Capture Polyhedron}

This subsection introduces the hypercube in normalized objective space and formalizes its role as the Pareto Capture Polyhedron. This construction is an invariant property of the multiobjective optimization problem and is therefore independent of any particular Pareto-generation method.

\vspace{0.8em}
\noindent
\fbox{%
\parbox{0.96\linewidth}{%
\textbf{Definition D1 --- Admissible Point}

 A point in normalized objective space is said to be \textit{admissible for Pareto capture}, or simply \textit{admissible}, if it satisfies \autoref{eq:NormObjCom}.
}}
\vspace{0.8em}

\textit{Every Pareto point is an admissible point; however, not every admissible point is Pareto optimal}. The admissible set is therefore defined by 
\begin{linenomath}
\begin{equation}
\label{eq:hypercube_First}
\mathcal{C}_n
=
\left\{
 \bar{\mu}^c\in\mathbb{R}^n
\;\middle|\;
0\leq \bar{\mu}_i^c \leq 1,
\;
i=1,\ldots,n
\right\}
\end{equation}
\end{linenomath}

\noindent where the superscript $c$ denotes a point in the hypercube. Thus, \autoref{eq:hypercube_First} defines the $n$-dimensional hypercube in normalized objective space.

We now formally define the notion of \textit{capture}.

\vspace{0.8em}
\noindent
\fbox{%
\parbox{0.96\linewidth}{%
\textbf{Definition D2 --- "Capture" or "Capturing" a Point}

A method is said to \textit{capture} an admissible point if it is structurally capable of generating that point as a solution.
}}
\vspace{0.8em}

\begin{figure}[htbp]
    \centering
    \includegraphics[width=0.94\textwidth]{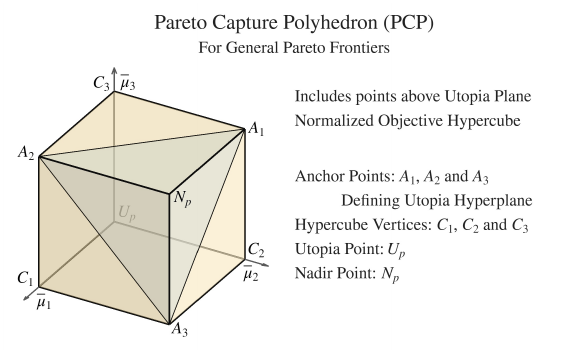}
    \vspace{-0.05in}
    \caption{The Pareto Capture Polyhedron is the hypercube in normalized objective space.}
    \label{fig:Pareto_Capture_polyhedron_First}
\end{figure}

 \autoref{eq:hypercube_First} naturally leads to the concept of the Pareto Capture Polyhedron in $n$-dimensional normalized objective space (see \autoref{fig:Pareto_Capture_polyhedron_First}). Among the vertices of the hypercube, the coordinate-axis vertices $C_1, C_2,\ldots, C_n$ are the standard basis vectors $e_1, e_2,\ldots, e_n$, where $e_i$ has value one in its $i$th coordinate and zero elsewhere. While \autoref{fig:Pareto_Capture_polyhedron_First} provides a geometric representation of the Pareto Capture Polyhedron, the following definition states the concept formally.

\vspace{0.8em}
\noindent
\fbox{
\parbox{0.96\linewidth}{%
\textbf{Definition D3 --- Pareto Capture Polyhedron}

A point belongs to the Pareto Capture Polyhedron if and only if it is admissible for Pareto capture. }}
\vspace{0.2cm}

\begin{figure}[htbp]
    \centering
    \includegraphics[width=0.94\textwidth]{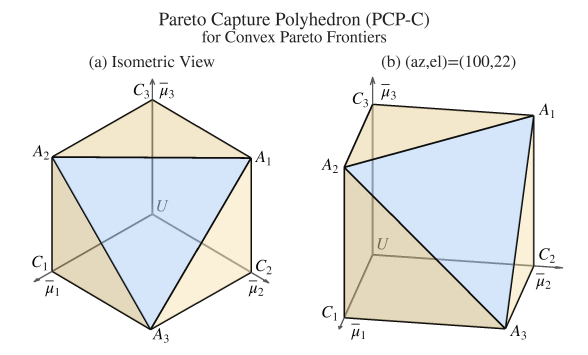}
    \vspace{-0.05in}
    \caption{Pareto Capture Polyhedron for convex Pareto frontiers in normalized objective space.}
\label{fig:Pareto_Capture_polyhedron_Convex}
\end{figure}

The darker shaded pyramidal region shown in \autoref{fig:Pareto_Capture_polyhedron_First} is an integral part of the general Pareto Capture Polyhedron. For a convex Pareto frontier in a multiobjective minimization problem, however, the admissible capture domain can be reduced by excluding this region. The resulting structure is defined as the Pareto Capture Polyhedron for Convex Pareto Frontiers (PCP-C) and is illustrated in \autoref{fig:Pareto_Capture_polyhedron_Convex}. 

The Pareto Capture Polyhedron defines the complete admissible region from which Pareto solutions may arise. Subsequent developments construct a generation grid on the Utopia Hyperplane and associate its points with locations in this polyhedron. Sections 3 and 4 then show that the restricted NNC generation domain captures only part of the admissible region, whereas GNC exploits the complete admissible generation domain.

We next introduce the Utopia Hyperplane, followed by the Pareto Capture Polygon and the corresponding Pareto Capture Grid used for Pareto frontier generation. 

\vspace{-0.3 cm}


\subsection{Utopia Hyperplane in \textit{n}-dimensional Normalized Objective Space}

We define the Utopia Hyperplane in \textit{n}-dimensional normalized objective space as follows.

\vspace{0.8em}
\noindent
\fbox{%
\parbox{0.96\linewidth}{%
\textbf{Definition D4 --- Utopia Hyperplane}

A \emph{Utopia Hyperplane} is an \textit{(n-1)}-dimensional hyperplane in the \textit{n}-dimensional Normalized Objective Space that contains the \textit{n} anchor points of the multiobjective optimization problem -- where \textit{n} is the number of objectives.
}}
\vspace{0.8em}
\vspace{0.8em}

\noindent While the Utopia Hyperplane plays an important role in the development of the NNC and GNC methods, its definition is invariant to any method under consideration. To express its equation, we first write the $k_{th}$ anchor point, for which all normalized objectives are equal to one except for the $k_{th}$ coordinate. Specifically, we have
\begin{linenomath}
\begin{equation}
A_k
=
\left(1,\ldots,1,0,1,\ldots,1\right)^T 
\end{equation}
\end{linenomath}
Since all anchor points belong, by definition, to the Utopia Hyperplane, its equation in normalized objective space is
\begin{linenomath}
\begin{equation}
\sum_{i=1}^{n}\bar{\mu}_i^p=n-1
\end{equation}
\end{linenomath}
or, equivalently
\begin{linenomath}
\begin{equation}
\label{eq:utopia_hyperplane}
\mathcal{H}_{\!p}
=
\left\{
\bar{\mu}^p \in \mathbb{R}^n
\;\middle|\;
\sum_{i=1}^{n} \bar{\mu}_i^p = n-1
\right\}
\end{equation}
\end{linenomath}
\noindent where the superscript \textit{p} stands for \textit{plane}, indicating that these coordinates represent points that lie on the Utopia Hyperplane. Direct substitution of the anchor points into the above equation points to its correctness. We note that the Pareto Capture Polyhedron is an $n$-dimensional object, whereas the Pareto Capture Polygon is an $(n-1)$-dimensional object.


\subsection{Pareto Capture Projection}

This subsection (i) defines the Pareto Capture Projection, (ii) motivates its introduction, and (iii) derives its mathematical formulation.\\

The Pareto Capture Projection is a normal projection that we formally define as follows.

\vspace{0.8em}
\noindent
\fbox{%
\parbox{0.96\linewidth}{%
\textbf{Definition D5 --- Pareto Capture Projection}

A \textit{Pareto Capture Projection} is a projection normal to the Utopia Hyperplane that associates a unique point on the Utopia Hyperplane with one or more points belonging to the Pareto Capture Polyhedron.
}}
\vspace{0.8em}

\noindent \textbf{Motivation}

The Pareto Capture Projection establishes a mapping between the Utopia Hyperplane and the Pareto Capture Polyhedron. It plays different roles in different methods. As shown in later sections, it plays distinct roles in the NNC and GNC methods. The mapping in question is ultimately used to help generate the Pareto frontier.\\

\noindent \textbf{Mathematical Derivation of the Pareto Capture Projection}

The Pareto Capture Polyhedron is defined in  \autoref{eq:hypercube_First} as,

\begin{linenomath}
\begin{equation}
\mathcal{C}_n
=
\left\{
\bar{\mu}^{\,c}\in\mathbb{R}^{n}
\;\middle|\;
0\leq \bar{\mu}^{\,c}_{i}\leq 1,\quad i=1,\ldots,n
\right\}
\end{equation}
\end{linenomath}
where the superscript $c$ denotes a point in the hypercube. The Utopia Hyperplane is defined by the normalized anchor points and is given by \autoref{eq:utopia_hyperplane} as
\begin{linenomath}
\begin{equation}
\mathcal{H}_{\!p}
=
\left\{
\bar{\mu}^{\,p}\in\mathbb{R}^{n}
\;\middle|\;
\sum_{i=1}^{n}\bar{\mu}^{\,p}_{i}=n-1
\right\}
\end{equation}
\end{linenomath}
where the superscript $p$ denotes a point on the Utopia Hyperplane.

Consider an arbitrary admissible point $\bar{\mu}^{\,c}\in\mathcal{C}_n$. The orthogonal projection of this point onto $\mathcal{H}_{\!p}$ is taken along the normal direction of the hyperplane. Since a normal vector to the Utopia Hyperplane is proportional to
\begin{linenomath}
\begin{equation}
E=(1,1,\ldots,1)^T
\end{equation}
\end{linenomath}

\noindent the projected point may be written as
\begin{linenomath}
\begin{equation}
\bar{\mu}^{\,p}
=
\bar{\mu}^{\,c}
+
\lambda
E
\end{equation}
\end{linenomath}
where $\lambda$ is a scalar.

Because $\bar{\boldsymbol{\mu}}^{\,p}$ lies on $\mathcal{H}_p$, it must satisfy
\begin{linenomath}
\begin{equation}
\sum_{i=1}^{n}\bar{\mu}^{\,p}_{i}=n-1
\end{equation}
\end{linenomath}
Substituting the projection expression yields
\begin{linenomath}
\begin{equation}
\sum_{i=1}^{n}
\left(
\bar{\mu}^{\,c}_{i}+\lambda
\right)
=
n-1
\end{equation}
\end{linenomath}
Therefore,
\begin{linenomath}
\begin{equation}
\sum_{i=1}^{n}\bar{\mu}^{\,c}_{i}
+
n\lambda
=
n-1
\end{equation}
\end{linenomath}
which yields
\begin{linenomath}
\begin{equation}
\lambda
=
\frac{n-1-\sum_{i=1}^{n}\bar{\mu}^{\,c}_{i}}{n}
\end{equation}
\end{linenomath}

Hence, the orthogonal projection from the Pareto Capture Polyhedron to the Utopia Hyperplane is

\begin{linenomath}
\begin{equation}
\bar{\mu}^{\,p}
=
\bar\mu^{\,c}
+
\frac{n-1-\sum_{i=1}^{n}\bar{\mu}^{\,c}_{i}}{n}
E
\end{equation}
\end{linenomath}

\noindent or equivalently,

\begin{linenomath}
\begin{equation}
\bar\mu^{\,p}
=
\bar{\mu}^{\,c}
+
\frac{n-1- E^T\bar{\mu}^{\,c}}{n}
E
\end{equation}
\end{linenomath}

\noindent or component-wise,
\begin{linenomath}
\begin{equation}
\bar{\mu}^{\,p}_{i}
=
\bar{\mu}^{\,c}_{i}
+
\frac{n-1-E^T\bar{\mu}^{\,c}}{n},
\qquad
i=1,\ldots,n
\end{equation}
\end{linenomath}

For the tri-objective case, $n=3$, the Utopia Hyperplane is
\begin{linenomath}
\begin{equation}
\bar{\mu}^{\,p}_{1}
+
\bar{\mu}^{\,p}_{2}
+
\bar{\mu}^{\,p}_{3}
=
2
\end{equation}
\end{linenomath}
and the projection becomes
\begin{linenomath}
\begin{equation}
\bar{\boldsymbol{\mu}}^{\,p}
=
\bar{\boldsymbol{\mu}}^{\,c}
+
\frac{2-\left(
\bar{\mu}^{\,c}_{1}
+
\bar{\mu}^{\,c}_{2}
+
\bar{\mu}^{\,c}_{3}
\right)}{3}
E
\end{equation}
\end{linenomath}

This projection maps the Pareto Capture Polyhedron onto the Pareto Capture Polygon on the Utopia Hyperplane. For three objectives, the projected image is a hexagon. More generally, for an $n$-objective problem, it is the projection of the Pareto Capture Polyhedron onto the $(n-1)$-dimensional Utopia Hyperplane.


\subsection{Pareto Capture Polygon and Pareto Capture Grid}

The \textit{Pareto Capture Polygon} lies on the Utopia Hyperplane. It is a geometric construction that serves two functions in the generic solution of the multiobjective optimization problem. First, it facilitates the construction of a grid within the Pareto Capture Polygon. We call this grid the \textit{Pareto Capture Grid} or the \textit{Grid}. Second, it will be used to construct the boundary of the Pareto Capture Polygon. The resulting Pareto Capture Grid also provides additional numerical efficiencies during Pareto frontier generation. The Pareto Capture Projection therefore plays a central role in establishing the geometric framework required for the complete capture of admissible points, without yet specifying the particular Pareto-generation method. We note again that the originating points for the projections are fully confined within the volumetric space of the \textit{Pareto Capture Polyhedron}. 

Every Pareto Capture Projection establishes a correspondence between a unique point in the Pareto Capture Grid and at least one corresponding point in the Pareto Capture Polyhedron. As will be seen, the construction process results in the \textit{complete capture} of the admissible Pareto frontier. 

We note that this development takes place in $n$-dimensional normalized objective space. Strictly speaking, the corresponding object is an $(n-1)$-dimensional polytope rather than a polygon. However, for the sake of familiarity, we will continue to use the term Polygon instead.\\

\noindent \textbf{Geometric Understanding of the Pareto Capture Polygon}

\begin{figure}[htbp]
    \centering
    \includegraphics[width=0.94\textwidth]{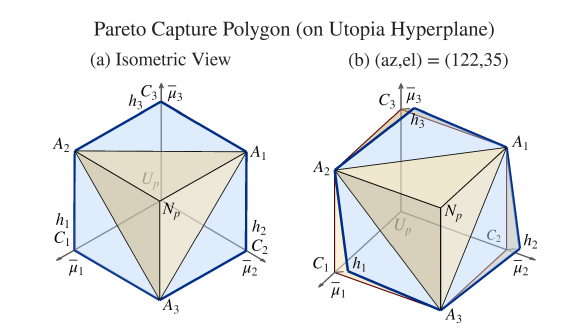}
    \vspace{-0.15in}
    \caption{Pareto Capture Polygon, which lies on the Utopia Hyperplane (a) Isometric view, (b) (az, el) = (122, 35)}
    \label{fig:Pareto_Capture_POLYGON}
\end{figure}

\autoref{fig:Pareto_Capture_POLYGON} provides a geometric interpretation of the Pareto Capture Polygon. Consider the orthogonal projection of a large number of points inside the Pareto Capture Polygon onto the Utopia Hyperplane. In doing so, we observe that all the points inside the Pareto Capture Polyhedron (shown in light tan) are projected onto the Utopia Hyperplane and intersect with the Utopia Hyperplane in light blue. These projected points form a hexagon on the Utopia Hyperplane --- which then forms the Pareto Capture Polygon. In three dimensions, this hexagon is the Pareto Capture Polygon.

The left panel is an isometric view of the Pareto Capture Polygon. The Anchors are denoted by $A_1$, $A_2$ and $A_3$. The right panel shows the same geometric objects, where the perspective is slightly changed.  It is seen that the outer boundaries of the hypercube are projected onto the Utopia Hyperplane to create a hexagon, in this three dimensional case. 

The same geometric result is now derived mathematically and is subsequently verified through numerical examples. In the process, the complete admissible Pareto frontier is generated.

\noindent\textbf{Mathematical derivation of the Pareto Capture Polygon Boundaries}

As previously stated, the Pareto Capture Polygon is the image of the Pareto Capture Polyhedron under orthogonal projection onto the Utopia Hyperplane. This Polygon is therefore the collection of all points $\bar{\mu}^{\,p}$ lying on the hyperplane, which are the projections of the complete set of hypercube points $\bar{\mu}^{\,c}$. These projected points satisfy the following Utopia Hyperplane equation
\begin{linenomath}
\begin{equation}
\sum_{i=1}^{n}\bar{\mu}^{\,p}_{i}=n-1 
\end{equation}
\end{linenomath}

To determine its boundary, recall that every point $\bar{\mu}^{\,p}$ on the Pareto Capture Polygon is obtained by projecting the corresponding point $\bar{\mu}^{\,c}$ in the normalized objective-space hypercube, where
\begin{linenomath}
\begin{equation}
0\leq \bar{\mu}^{\,c}_{i}\leq 1,
\qquad i=1,\ldots,n 
\end{equation}
\end{linenomath}

The projection from the hypercube to the Utopia Hyperplane is made in the direction $\left(1,\ldots,1\right)^T$. Consequently, all points on a common \textit{\textbf{projection fiber}} differ only by an additive scalar in every coordinate. Consequently, a projected point $\bar{\mu}^{\,p}$ belongs to the Pareto Capture Polygon if and only if there exists a scalar $\lambda$ such that
\begin{linenomath}
\begin{equation}
\bar{\mu}^{\,c}_{i}
=
\bar{\mu}^{\,p}_{i}
-
\lambda,
\qquad i=1,\ldots,n 
\end{equation}
\end{linenomath}
with
\begin{linenomath}
\begin{equation}
0\leq
\bar{\mu}^{\,p}_{i}
-
\lambda
\leq 1,
\qquad i=1,\ldots,n 
\end{equation}
\end{linenomath}

The above inequalities imply
\begin{linenomath}
\begin{equation}
\bar{\mu}^{\,p}_{i}-1
\leq
\lambda
\leq
\bar{\mu}^{\,p}_{i}
\qquad i=1,\ldots,n 
\end{equation}
\end{linenomath}

Therefore, such a scalar $\lambda$ exists if and only if the intervals
$\left[\bar{\mu}^{\,p}_{i}-1,\bar{\mu}^{\,p}_{i}\right]$ have a nonempty common intersection. This is equivalent to
\begin{linenomath}
\begin{equation}
\max_{i}
\left(\bar{\mu}^{\,p}_{i}-1\right)
\leq
\min_{i}
\bar{\mu}^{\,p}_{i}
\end{equation}
\end{linenomath}

Equivalently,
\begin{linenomath}
\begin{equation}
\max_{i}\bar{\mu}^{\,p}_{i}
-
\min_{i}\bar{\mu}^{\,p}_{i}
\leq
1 
\end{equation}
\end{linenomath}

Thus, the Pareto Capture Polygon is the set
\begin{linenomath}
\begin{equation}
\mathcal{P}_{\mathrm{PC}}
=
\left\{
\bar{\mu}^{\,p}\in\mathbb{R}^{n}
\;\middle|\;
\sum_{i=1}^{n}\bar{\mu}^{\,p}_{i}=n-1,
\quad
\max_{i}\bar{\mu}^{\,p}_{i}
-
\min_{i}\bar{\mu}^{\,p}_{i}
\leq
1
\right\}
\end{equation}
\end{linenomath}

The boundary of the Pareto Capture Polygon is obtained when the above inequality is active:
\begin{linenomath}
\begin{equation}
\max_{i}\bar{\mu}^{\,p}_{i}
-
\min_{i}\bar{\mu}^{\,p}_{i}
=
1 
\end{equation}
\end{linenomath}

Equivalently, the boundary consists of all points on the Utopia Hyperplane for which at least one pair of coordinates differs by exactly one:
\begin{linenomath}
\begin{equation}
\bar{\mu}^{\,p}_{i}
-
\bar{\mu}^{\,p}_{j}
=
1,
\qquad
i\neq j 
\end{equation}
\end{linenomath}

Hence, the boundary of the Pareto Capture Polygon may be written as the union
\begin{linenomath}
\begin{equation}
\partial \mathcal{P}_{\mathrm{PC}}
=
\bigcup_{i\neq j}
\left\{
\bar{\mu}^{\,p}\in\mathbb{R}^{n}
\;\middle|\;
\sum_{k=1}^{n}\bar{\mu}^{\,p}_{k}=n-1,
\quad
\bar{\mu}^{\,p}_{i}-\bar{\mu}^{\,p}_{j}=1,
\quad
\max_{k}\bar{\mu}^{\,p}_{k}
-
\min_{k}\bar{\mu}^{\,p}_{k}
\leq 1
\right\}
\end{equation}
\end{linenomath}

For the tri-objective case, the Utopia Hyperplane is
\begin{linenomath}
\begin{equation}
\bar{\mu}^{\,p}_{1}
+
\bar{\mu}^{\,p}_{2}
+
\bar{\mu}^{\,p}_{3}
=
2 
\end{equation}
\end{linenomath}

The boundary conditions are
\begin{linenomath}
\begin{equation}
\bar{\mu}^{\,p}_{i}
-
\bar{\mu}^{\,p}_{j}
=
1,
\qquad i\neq j 
\end{equation}
\end{linenomath}
which generate the six sides of the hexagon. Therefore, for three objectives, the Pareto Capture Polygon is the hexagon obtained by orthogonally projecting the Pareto Capture Polyhedron onto the Utopia Hyperplane.\\


\subsection{Mathematical Representation of Pareto Capture Grid }

The Pareto Capture Grid is obtained by projecting the normalized objective-space hypercube onto the Utopia Hyperplane. The purpose of this section is to describe the resulting projected region using linear inequalities.

Let the normalized objective-space hypercube be

\begin{linenomath}
\begin{equation}
\mathcal{C}_{n}
=
\left\{
\bar{\mu}^{\,c}\in\mathbb{R}^{n}
\;\middle|\;
0\leq \bar{\mu}^{\,c}_{i}\leq 1,
\quad i=1,\ldots,n
\right\}.
\end{equation}
\end{linenomath}

The Pareto Capture Projection maps a point $\bar{\mu}^{,c}$ belonging to the Pareto Capture Polyhedron to a point $\bar{\mu}^{\,p}$ on the Utopia Hyperplane. The Utopia Hyperplane is

\begin{linenomath}
\begin{equation}
\sum_{i=1}^{n}\bar{\mu}^{\,p}_{i}=n-1.
\end{equation}
\end{linenomath}

The orthogonal projection from $\bar{\mu}^{\,c}$ to $\bar{\mu}^{\,p}$ is

\begin{linenomath}
\begin{equation}
\bar{\mu}^{\,p}_{i}
=
\bar{\mu}^{\,c}_{i}
+
\frac{
n-1-\displaystyle\sum_{j=1}^{n}\bar{\mu}^{\,c}_{j}
}{n},
\qquad
i=1,\ldots,n.
\end{equation}
\end{linenomath}

To characterize the Pareto Capture Grid analytically, consider any subset $S$ of the objective indices. For example, if $n=5$, one possible subset is $S=\{1,3,5\}$. The number of elements in this subset is denoted by

\begin{linenomath}
\begin{equation}
m=|S|.
\end{equation}
\end{linenomath}

We now examine the sum of the projected coordinates associated with the indices in $S$:

\begin{linenomath}
\begin{equation}
\sum_{i\in S}\bar{\mu}^{\,p}_{i}.
\end{equation}
\end{linenomath}

Using the projection formula, we obtain

\begin{linenomath}
\begin{equation}
\sum_{i\in S}\bar{\mu}^{\,p}_{i}
=
\sum_{i\in S}\bar{\mu}^{\,c}_{i}
+
\frac{m}{n}
\left(
n-1-\sum_{j=1}^{n}\bar{\mu}^{\,c}_{j}
\right).
\end{equation}
\end{linenomath}

The minimum value of this quantity occurs when the coordinates in $S$ are as small as possible and the coordinates outside $S$ are as large as possible. Thus,

\begin{linenomath}
\begin{equation}
\bar{\mu}^{\,c}_{i}=0
\quad \mathrm{for}\quad i\in S,
\qquad
\bar{\mu}^{\,c}_{i}=1
\quad \mathrm{for}\quad i\notin S.
\end{equation}
\end{linenomath}

Since there are $n-m$ coordinates outside $S$,

\begin{linenomath}
\begin{equation}
\sum_{j=1}^{n}\bar{\mu}^{\,c}_{j}=n-m.
\end{equation}
\end{linenomath}

Substitution gives

\begin{linenomath}
\begin{equation}
\sum_{i\in S}\bar{\mu}^{\,p}_{i}
=
\frac{m}{n}
\left(
n-1-(n-m)
\right)
=
\frac{m(m-1)}{n}.
\end{equation}
\end{linenomath}

The maximum value occurs when the coordinates in $S$ are as large as possible and the coordinates outside $S$ are as small as possible. Thus,

\begin{linenomath}
\begin{equation}
\bar{\mu}^{\,c}_{i}=1
\quad \mathrm{for}\quad i\in S,
\qquad
\bar{\mu}^{\,c}_{i}=0
\quad \mathrm{for}\quad i\notin S.
\end{equation}
\end{linenomath}

Therefore,

\begin{linenomath}
\begin{equation}
\sum_{i\in S}\bar{\mu}^{\,c}_{i}=m,
\qquad
\sum_{j=1}^{n}\bar{\mu}^{\,c}_{j}=m.
\end{equation}
\end{linenomath}

Substitution gives

\begin{linenomath}
\begin{equation}
\sum_{i\in S}\bar{\mu}^{\,p}_{i}
=
m
+
\frac{m}{n}
\left(
n-1-m
\right)
=
\frac{m(2n-1-m)}{n}.
\end{equation}
\end{linenomath}

Therefore, for every subset $S\subset\{1,2,\ldots,n\}$, with $m=|S|$, the projected hypercube satisfies

\begin{linenomath}
\begin{equation}
\frac{m(m-1)}{n}
\leq
\sum_{i\in S}\bar{\mu}^{\,p}_{i}
\leq
\frac{m(2n-1-m)}{n}.
\end{equation}
\end{linenomath}

Together with the Utopia Hyperplane equation, 

\begin{linenomath}
\begin{equation}
\sum_{i=1}^{n}\bar{\mu}^{\,p}_{i}=n-1,
\end{equation}
\end{linenomath}

\noindent \textbf{\textit{these inequalities define the Pareto Capture Grid}}.

As a simple special case, take $m=1$. Then the subset $S$ contains only one coordinate, and the inequality becomes

\begin{linenomath}
\begin{equation}
0
\leq
\bar{\mu}^{\,p}_{i}
\leq
2-\frac{2}{n},
\qquad
i=1,\ldots,n.
\end{equation}
\end{linenomath}

For $n=3$, this gives

\begin{linenomath}
\begin{equation}
0
\leq
\bar{\mu}^{\,p}_{i}
\leq
\frac{4}{3},
\qquad
i=1,2,3.
\end{equation}
\end{linenomath}

Combining these coordinate bounds with

\begin{linenomath}
\begin{equation}
\bar{\mu}^{\,p}_{1}
+
\bar{\mu}^{\,p}_{2}
+
\bar{\mu}^{\,p}_{3}
=
2
\end{equation}
\end{linenomath}

\textit{\textbf{produces the Pareto Capture Polygon}}.


\subsection{Projection-Line Collapse and the Minimal Representative Set on Pareto Capture Polygon}

In the process of generating a distributed set of points within the volume of the Pareto Capture Polyhedron, a geometric structure emerges that has important computational implications, as discussed below. While it was within the realm of possibilities that the cost of guaranteeing the capture of the complete set of admissible points might entail a significant computational cost, the geometric and mathematical developments presented in this section show that this is not the case. \\

\textbf{Geometric Interpretation of Projection-Line (Projection-Fiber) Collapse}
\vspace{12pt}

\begin{figure}[htbp]
    \centering
    \includegraphics[width=0.94\textwidth]{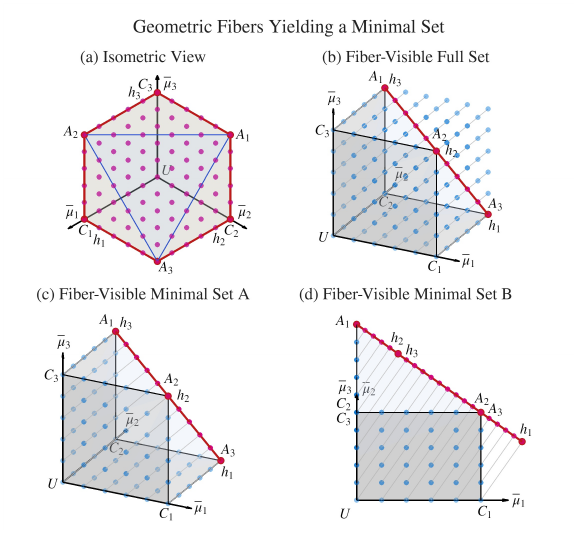}
    \vspace{-0.05in}
    \caption{Geometric fibers yield a minimal set of points. Vector Views are (a) (1,1,1), (b) (1,-2,1), (c) (0,-1,1), and (d) (1,-2,1).}
    \label{fig:from_nnc_to_gnc_projection}
\end{figure}

\autoref{fig:from_nnc_to_gnc_projection} illustrates important properties of an evenly distributed set of points within the volume of the Pareto Capture Polyhedron --- in normalized objective space. This figure illustrates the Pareto Capture Polyhedron containing an evenly distributed set of points (shown in blue). In addition, the Pareto Capture Polygon is displayed (in red). The selected viewing directions reveal several important geometric properties.

The Anchor points are denoted by $A_1$ $A_2$ and $A_3$. The vertices of the hypercube on the axes are seen as $C_1$ $C_2$ and $C_3$. The utopia point is also shown, $U$.

Subplot (a) shows an isometric view (a view vector (1,1,1). With this view, a series of points are behind each other, thereby revealing only a subset of the admissible points. The points in the Pareto Capture Polyhedron are projected onto the Pareto Capture Polygon as a hexagon. 

Subplot (b) shows a perspective from the view vector (1,-2,1). From this view, the Polygon collapses into a line. It is observed that a series of points lie on the same  projection line/fiber, which is normal to the Pareto Capture Polygon.

This observation immediately implies that, since multiple admissible points project to the same point onto the Pareto Capture Polygon, it is only necessary to use one of these admissible points. Accordingly, only one representative point is required from each projection line. In this work, the representative point is chosen to lie on the \textit{zero-coordinate faces}; namely, the ones for which $\bar{\mu}_1=0$, $\bar{\mu}_2=0$, and $\bar{\mu}_3=0$. The following subplots illustrate this choice.

In subplot (c), the view vector is (0,-1,1). As can be seen, multiplicity is avoided by only keeping the admissible points on the zero-coordinate surfaces. These points --- alone --- constitute the complete representative set of admissible points of the Pareto Capture Polyhedron. These points provide a minimal representative set of complete admissible points on the Pareto Capture Polyhedron --- without loss of information. Similar behavior is observed in Subplot (d), this time using view vector (1,-2,1).

\vspace{12pt}
\textbf{Mathematical Proof of Projected Fiber Collapse and Resulting Advantages}
\vspace{12pt}

A natural approach for generating the Pareto Capture Polygon is to begin with a structured distribution of points throughout the Pareto Capture Polyhedron. Let the Pareto Capture Polyhedron be the normalized objective-space hypercube

\begin{linenomath}
\begin{equation}
\mathcal{C}_{n}
=
\left\{
\bar{\mu}^{\,c}\in\mathbb{R}^{n}
\;\middle|\;
0\leq\bar{\mu}^{\,c}_{i}\leq1,
\quad
i=1,\ldots,n
\right\}.
\end{equation}
\end{linenomath}

Assume that each coordinate direction is divided uniformly into $n_d$ intervals, producing the volumetric lattice

\begin{linenomath}
\begin{equation}
\bar{\mu}^{\,c}
=
\frac{1}{n_d}
\left(
i_1,i_2,\ldots,i_n
\right)^T,
\qquad
i_k\in\{0,1,\ldots,n_d\}.
\end{equation}
\end{linenomath}

The orthogonal projection of every lattice point onto the Utopia Hyperplane,

\begin{linenomath}
\begin{equation}
\sum_{i=1}^{n}\bar{\mu}^{\,p}_{i}=n-1,
\end{equation}
\end{linenomath}

is performed along the normal direction

\begin{linenomath}
\begin{equation}
E=(1,1,\ldots,1)^T.
\end{equation}
\end{linenomath}

Accordingly, the projection of an arbitrary point is

\begin{linenomath}
\begin{equation}
\bar{\mu}^{\,p}
=
\bar{\mu}^{\,c}
+
\frac{
n-1-\displaystyle\sum_{i=1}^{n}\bar{\mu}^{\,c}_{i}
}{n}
E.
\end{equation}
\end{linenomath}

Now consider two lattice points related by

\begin{linenomath}
\begin{equation}
\bar{\mu}^{\,c,b}
=
\bar{\mu}^{\,c,a}
+
\gamma E,
\end{equation}
\end{linenomath}

where $\gamma$ is an arbitrary scalar. Their projections satisfy

\begin{linenomath}
\begin{equation}
\bar{\mu}^{\,p,b}
=
\bar{\mu}^{\,c,a}
+
\gamma E
+
\frac{
n-1-\displaystyle\sum_{i=1}^{n}\bar{\mu}^{\,c,a}_{i}
-n\gamma
}{n}
E.
\end{equation}
\end{linenomath}

Since the terms containing $\gamma$ cancel identically,

\begin{linenomath}
\begin{equation}
\bar{\mu}^{\,p,b}
=
\bar{\mu}^{\,p,a}.
\end{equation}
\end{linenomath}

This result shows that every collection of lattice points lying on a common line parallel to $E$ projects onto exactly the same point of the Pareto Capture Polygon. Thus, although the volumetric lattice contains many distinct points, only one representative point is required from each such projection line.

A natural choice is the first lattice point encountered on the zero-coordinate faces of the Pareto Capture Polyhedron. This representative set is characterized by

\begin{linenomath}
\begin{equation}
\boxed{
\min(i_1,i_2,\ldots,i_n)=0.
}
\end{equation}
\end{linenomath}

For the tri-objective case,

\begin{linenomath}
\begin{equation}
\boxed{
\min(i,j,k)=0.
}
\end{equation}
\end{linenomath}
\vspace{0.1cm}

Consequently, the complete Pareto Capture Polygon is generated by projecting only this zero-face representative set, rather than the entire volumetric lattice.\\

\vspace{0.15cm}
\noindent
\fbox{%
\parbox{0.96\linewidth}{%
\textbf{Key Result.}
The zero-face representative set defined by
$\min(i_1,i_2,\ldots,i_n)=0$
contains exactly one representative from every projection line (or, projection fiber). Consequently, it generates exactly the same Pareto Capture Polygon as the complete volumetric lattice while reducing the projection stage from
\vspace{-4pt}
\[
\begin{array}{c}
\text{Order}\; O\!\left(n_d^{\,n}\right)\\[2pt]
\hspace{-7cm} \text{to}\\[2pt]
\text{Order}\; O\!\left(n_d^{\,n-1}\right)
\end{array}
\]
\vspace{2pt}
Thus, the \textit{complete} Pareto Capture Polygon and Pareto Capture Grid are obtained without loss of information while reducing the projection computation by one order in dimensionality.
\vspace{2pt}
}%
}

  
\subsection{Method Independence of the above Definitions and Properties}

 At this point, it is useful to emphasize that the developments presented in this section reveal inherent geometric properties of the multiobjective optimization problem. These geometric properties are independent of any particular Pareto-generation method, including NNC, GNC, Weighted Sum, and Compromise Programming. Ideally, any Pareto-generation method should be capable of capturing every admissible point within the Pareto Capture Polyhedron. Several existing methods, however, are structurally unable to satisfy this key objective, e.g., the Weighted Sum method \citep{Koski1985}, NNC, and Normal Boundary Intersection (NBI). The structural reasons for these limitations are discussed in the following sections. The distinct feature of GNC is that it has the ability to capture the \textit{complete} admissible Pareto frontier.\\
 
 Consequently, the ability to capture all admissible points provides a natural geometric criterion for evaluating Pareto-generation methods.\\

\textit{Every aspect of the GNC methodology follows directly from these invariant geometric structures. Accordingly, the proposed GNC method is not introduced as an ad hoc algorithmic construction, but instead emerges naturally from the invariant geometry of the multiobjective optimization problem.}\\

\textit{ Collectively, the geometric structures developed in this section provide a common framework for defining, analyzing, comparing, and extending Pareto-generation methods.}


\section{The Normalized Normal Constraint Method --- Strengths and Limitations}

The purpose of this section is not to reformulate the NNC method in detail, but rather to identify the specific geometric characteristics that distinguish it from the proposed GNC framework. The Normalized Normal Constraint (NNC) method begins with the same multiobjective optimization problem, anchor-point evaluations, utopia point, nadir point, normalized objective space, and Utopia Hyperplane. Therefore, these elements are not repeated here. The purpose of this section is to summarize the NNC-specific grid construction and normal-constraint formulation needed for the subsequent comparison with GNC. 

Conceptually, NNC samples normals from a restricted triangular region of the utopia plane, thereby capturing only a subset of admissible points. GNC instead constructs the smallest geometric structure capable of generating normal projections that span the \textit{complete} admissible Pareto frontier. The resulting \textit{Pareto Capture Polygon} preserves the simplicity of NNC while removing its fundamental geometric incompleteness. In the Validation Section, a computational comparative perspective is provided.

This Section presents the following key comparative aspects of NNC and GNC in sequence:

\begin{itemize}
    \item NNC and GNC: Respective Exploitations of the Fundamental Geometric Structures
    \item NNC and GNC’s Respective Mathematical Treatments of the Hyperplane and Grid Structures
    \item Geometric View of NNC’s Pareto-Incompleteness 
    \item Quantitative View of NNC’s Pareto-Incompleteness
\end{itemize}
 
  
\subsection{NNC and GNC: Respective Exploitations of the Fundamental Geometric Structures}

Most developments in the multiobjective optimization literature begin with the question "Given the generic multiobjective problem (MOP), what is the best approach to generate the \textit{complete} admissible Pareto frontier?" The present work instead begins by asking a different question; "Before developing a new generation method or improving an existing one, what are the geometric objects and structures that are inherently induced by the normalized multiobjective problem/space itself?" Each of these questions does indeed belong to the MOP endeavor. Section~\ref{NNC_Incompleteness} shows how NNC is a partial utilization of these structures, while GNC is a complete utilization thereof.

We note some fundamental distinctions between NNC and GNC. NNC follows a traditional forward progression, from problem formulation and development to solution algorithm, while GNC employs a geometry-driven development. In addition, NNC constructs a simplex (a triangle in three dimensions) to create a grid, which is used for a series of normal projections, which are used to generate an \textit{incomplete} set of Pareto solutions, by its very structural construction. At no point is there an expectation that the generated Pareto set will be, in general, complete. In fact, it is later shown that for an $n$-objective problem, the fraction of \textit{admissible} points that NNC \textit{captures} may decrease factorially as $1/(n-1)!$; and the corresponding missed fraction may increase to $1-1/(n-1)!$. This is proven below. By contrast, GNC has the ability to generate 100\% of the admissible points.\\


\subsection{NNC and GNC's Respective Mathematical Treatments of the Hyperplane and Grid Structures}

As a fundamental departure from the NNC approach, GNC begins by identifying the complete set of \textit{admissible points}, which by definition reside in the \textit{Pareto Capture Polyhedron}. From there it generates the corresponding Pareto Capture Polygon (a hexagon for three objectives) that can be used directly to capture 100\% of the admissible points. For three objectives, this Polygon is a hexagon for GNC --- and not a triangle as in NNC. This geometric construction provides the ability for GNC to generate the \textit{complete} set of admissible points. Note also that the hexagon circumscribes the triangle, providing further evidence of the \textit{incomplete} nature of NNC.\\

\noindent\textit{It is important to note that, unlike the \textbf{Pareto Capture Grid} in this paper, which is generated directly in normalized objective space, the NNC grid is generated in barycentric coordinates}  $\alpha$. \textit{The corresponding points on the common Utopia Hyperplane are subsequently mapped back into normalized objective space during the solution process.}\\

The Utopia Hyperplane in normalized objective space is

\begin{linenomath}
\begin{equation}
\sum_{i=1}^{n}\bar{\mu}^{\,p}_{i}=n-1.
\end{equation}
\end{linenomath}

\noindent \textit{This hyperplane is identical to that used in the original NNC formulation. The difference lies only in the coordinate representation: the present paper uses normalized objective coordinates $\bar{\mu}$, whereas the original NNC formulation employs barycentric coordinates $\alpha$ on the NNC simplex.}\\

NNC constructs its grid on the simplex defined by the normalized anchor points. Let the normalized anchor points be denoted by $A_1,A_2,\ldots,A_n$, where $A_i$ is the anchor point associated with the objective $i$. A point in the NNC simplex is generated using barycentric coordinates $\alpha_i$ as

\begin{linenomath}
\begin{equation}
\bar{\mu}^{\,p}_{\mathrm{NNC}}
=
\sum_{i=1}^{n}\alpha_i A_i,
\end{equation}
\end{linenomath}

\noindent where

\begin{linenomath}
\begin{equation}
\sum_{i=1}^{n}\alpha_i=1,
\qquad
\alpha_i\geq 0,
\qquad
i=1,\ldots,n.
\end{equation}
\end{linenomath}

Thus, the NNC grid is generated by distributing the barycentric coordinates $\alpha_i$ over the simplex. For $n_d$ grid divisions, the admissible grid indices satisfy

\begin{linenomath}
\begin{equation}
i_1+i_2+\cdots+i_n=n_d,
\qquad
i_k\in\{0,1,\ldots,n_d\},
\end{equation}
\end{linenomath}

\noindent with

\begin{linenomath}
\begin{equation}
\alpha_k=\frac{i_k}{n_d},
\qquad
k=1,\ldots,n.
\end{equation}
\end{linenomath}

The resulting number of NNC grid points is

\begin{linenomath}
\begin{equation}
N_{\mathrm{NNC}}
=
\binom{n_d+n-1}{n-1}.
\end{equation}
\end{linenomath}

For each NNC grid point, a normal constraint subproblem is solved. Let $\bar{\mu}^{\,p}_{\mathrm{NNC},j}$ denote the $j$th NNC grid point on the Utopia Hyperplane. The normal direction is assumed normal to the Utopia Hyperplane, namely along the direction $E=(1,1,\ldots,1)^T$. The corresponding normal line may be expressed as

\begin{linenomath}
\begin{equation}
\bar{\mu}^{\,p}_{\mathrm{NNC},j}+tE,
\end{equation}
\end{linenomath}

\noindent where $t$ is a scalar line parameter.

The NNC subproblem associated with grid point $j$ may therefore be written in normalized objective space as

\begin{linenomath}
\begin{equation}
\min_{x}\; \bar{\mu}_{k}(x),
\end{equation}
\end{linenomath}

\noindent subject to the original design constraints and the normal constraints

\begin{linenomath}
\begin{equation}
\left(
\bar{\mu}(x)-\bar{\mu}^{\,p}_{\mathrm{NNC},j}
\right)^T
\left(
A_i-\bar{\mu}^{\,p}_{\mathrm{NNC},j}
\right)
\leq 0,
\qquad
i=1,\ldots,n-1.
\end{equation}
\end{linenomath}

Here, $\bar{\mu}(x)$ denotes the normalized objective vector associated with design point $x$, and the index $k$ denotes the objective selected for minimization in the corresponding objective order. The original behavioral constraints and side constraints remain unchanged from the multiobjective problem statement.

The preceding discussion completes the summary of the NNC formulation required for the present development. The preceding equations show that NNC uses the same normalized objective space and the same Utopia Hyperplane as GNC, but its grid coordinates $\alpha_i$ are barycentric coordinates on the NNC simplex. The following section examines the geometric implications of restricting the normal-generation grid to this simplex and compares this construction with the generalized framework developed in the present work.


\subsection{Geometric View of NNC's Pareto-Incompleteness}
\label{NNC_Incompleteness}

Over the past quarter century, NNC has been used successfully in a wide range of technical applications. During that time, a key challenge common to many deterministic Pareto-generation methods also affected NNC. Namely, the inability to capture the complete admissible Pareto frontier, which is fully enclosed in the Pareto Capture Polyhedron.

This important limitation of NNC arises because its grid for Pareto frontier generation is structurally incomplete. In three dimensions, the grid is confined to the NNC Triangle on the utopia plane. Consequently, only admissible Pareto regions intersected by normals emanating from this restricted grid can be captured. Pareto regions lying outside the NNC projection domain remain inaccessible, regardless of any increase in grid density or other numerical refinement. This is the case because this deficiency is  structural. Therefore, ideas such as increasing the density of the NNC grid do not help as they do not impact the structural incompleteness of the Pareto frontier construction process.

\begin{figure}[htbp]
    \centering
    \includegraphics[width=0.94\textwidth]{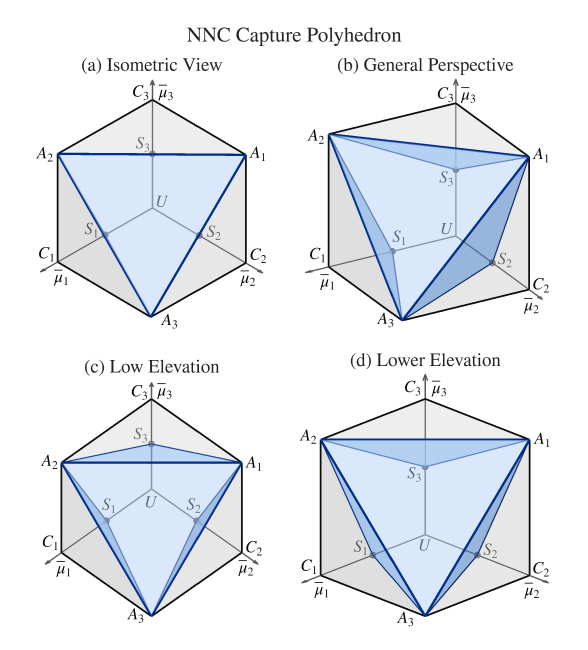}
    \vspace{-0.05in}
    \caption{The NNC Capture Polyhedron is shown from four complementary viewpoints. The isometric view establishes the canonical geometry, while the remaining views reveal the polyhedral side walls and volumetric structure associated with the NNC Capture Region. The (az,el) coordinates are (a) (1,1,1) (b) (150, 25), (c) (135, 45) and (d) (135, 23)}
    \label{fig:NNC_Capture_Polyhedron2}
\end{figure}

The NNC Pareto frontier incompleteness phenomenon can be clearly seen from a graphical perspective. \autoref{fig:NNC_Capture_Polyhedron2} displays the NNC limitation using four subplots of the NNC Capture Polyhedron, illustrated for a convex Pareto frontier to improve visual clarity. Note, however, that all of these observations remain valid regardless of convexity. 

These figures show the Pareto Capture Polyhedron in normalized objective space. $A_1$, $A_2$, and $A_3$ denote the anchor points. The blue regions represent the admissible points that can be captured by NNC. The gray regions represent the admissible points that cannot be captured by NNC. In sum, 50\% of the admissible points lie outside the admissible capture domain of NNC. The resulting geometric conclusion can be understood through careful examinations of the four subplots. Subplot (a) displays an isometric view where the normal projection is exactly below the boundary of the NNC simplex. As different views are chosen, the more complete view of the NNC solutions emerges.

\begin{figure}[htbp]
    \centering
    \includegraphics[width=0.94\textwidth]{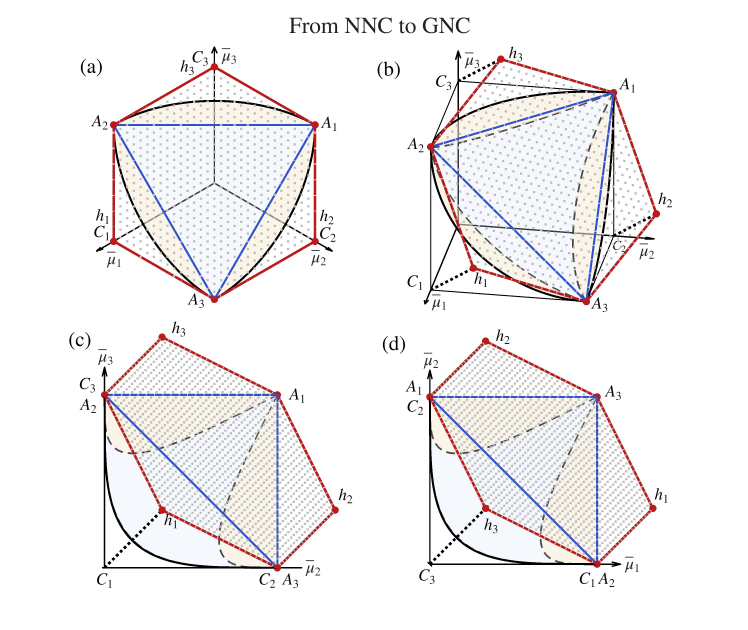}
    \vspace{-0.15in}
    \caption{Contrasting NNC's partial Pareto frontier capture to GNC's complete capture. (a) Isometric view, (b) (az, el) = (100, 25), (c) and (d) are orthogonal views.}
    \label{fig:from_nnc_to_gnc_overview}
\end{figure}

The geometric distinction between NNC and GNC becomes particularly evident in \autoref{fig:from_nnc_to_gnc_overview} where a generic Pareto frontier is included in the illustration, providing further insight into the respective geometric properties of NNC and GNC. 

We describe these illustrations as follows. The red hexagon represents the Pareto Capture Polygon, which is used in the GNC process. The blue triangle represents the Grid used by NNC, which is circumscribed by the hexagon. Both the triangle and the hexagon lie on the Utopia Hyperplane. The three black curves denote the boundaries of the Pareto frontier. 

The grid points within the NNC triangle serve as the origins of the normal projections that generate the corresponding Pareto solutions. Because the NNC triangle spans only a subset of the Pareto Capture Polygon, these normal projections are structurally capable of capturing only a subset of the admissible points.

In GNC, the NNC triangle is replaced by the Pareto Capture Polygon (a hexagon in the tri-objective case). Consequently, the corresponding normal projections originate from the complete Pareto Capture Polygon and are therefore capable of capturing the complete set of admissible points.

The particular angular perspectives are noted in the figure caption. Collectively, these views show that NNC provides \textit{partial capture} of the admissible Pareto frontier, whereas GNC provides \textit{complete capture}.

The above geometric result also applies directly to the Normal Boundary Intersection (NBI) method. Since NBI employs the same simplex-based projection structure as NNC, it structurally inherits the same admissible-point capture limitation established here.

\subsection{Quantitative View of NNC's Pareto-Incompleteness}

The preceding section established that the NNC method constructs its grid on the simplex defined by the normalized anchor points. This section examines the geometric implications of that construction.

The complete admissible projection domain for Pareto frontier generation is the Pareto Capture Polygon, defined by

\begin{linenomath}
\begin{equation}
\mathcal{P}_{\mathrm{PC}}
=
\left\{
\bar{\mu}^{\,p}\in\mathbb{R}^{n}
\;\middle|\;
\sum_{i=1}^{n}\bar{\mu}^{\,p}_{i}=n-1,
\;
\max_i\left(\bar{\mu}^{\,p}_{i}\right)
-
\min_i\left(\bar{\mu}^{\,p}_{i}\right)
\leq1
\right\}
\end{equation}
\end{linenomath}

However, by construction, the NNC grid is generated only over the simplex

\begin{linenomath}
\begin{equation}
\Delta_{\mathrm{NNC}}^{\,n-1}
=
\left\{
\bar{\mu}^{\,p}
=
\sum_{i=1}^{n}\alpha_iA_i
\;\middle|\;
\sum_{i=1}^{n}\alpha_i=1
\;
\alpha_i\ge0
\right\}
\end{equation}
\end{linenomath}

\noindent which is a subset of the Pareto Capture Polygon. Hence,

\begin{linenomath}
\begin{equation}
\Delta_{\mathrm{NNC}}^{\,n-1}
\subset
\mathcal{P}_{\mathrm{PC}},
\qquad n\ge3
\end{equation}
\end{linenomath}

Since every Pareto point generated by NNC must originate from a normal constructed at a point of
$\Delta_{\mathrm{NNC}}^{\,n-1}$,
the method is unable to generate Pareto points whose normal projections would have to originate outside this simplex.

Accordingly, the missed admissible region is

\begin{linenomath}
\begin{equation}
\mathcal{P}_{\mathrm{missed}}
=
\left\{
\bar{\mu}^{\,p}
\;\middle|\;
\bar{\mu}^{\,p}\in\mathcal{P}_{\mathrm{PC}}
\;\mathrm{and}\;
\bar{\mu}^{\,p}\notin
\Delta_{\mathrm{NNC}}^{\,n-1}
\right\}
\end{equation}
\end{linenomath}

Therefore,

\begin{linenomath}
\begin{equation}
\mathcal{P}_{\mathrm{missed}}
\neq
\emptyset,
\qquad n\ge3
\end{equation}
\end{linenomath}

\noindent which establishes that the NNC method is geometrically incapable of capturing the complete admissible Pareto region.

For the tri-objective case, the Pareto Capture Polygon is the regular hexagon developed in the preceding section, whereas the NNC generation domain is the inscribed equilateral triangle formed by the anchor points. 

Consequently, the three regions of the Pareto Capture Polygon (a hexagon in the tri-objective case) lying outside the NNC triangle consist of admissible points that are inaccessible to the NNC method. To quantify the respective performances of NNC and GNC, we show, then develop a closed-form expression of the  NNC admissible capture fraction as

\begin{linenomath}
\begin{equation}
\frac{A_{\mathrm{NNC}}}
     {A_{\mathrm{PC}}}
=
\frac{1}{(n-1)!}
\end{equation}
\end{linenomath}

\noindent which equals $1/2$ for three objectives and decreases factorially as the number of objectives increases. This equation is proved in this next section.\\

\textit{\textbf{Closed-Form Derivation of the NNC Capture Fraction}}\\

In the following, we derive the closed-form expression for the fraction of the admissible projection domain (on the zero-coordinate faces) that the NNC method misses.\\

\begin{linenomath}
\begin{equation}
\mathcal{C}_n=\left[0,1\right]^n 
\end{equation}
\end{linenomath}
The admissible distinct projected points lie on the zero-coordinate faces,
\begin{linenomath}
\begin{equation}
\partial_0 \mathcal{C}_n
=
\left\{
\boldsymbol{\mu}\in\left[0,1\right]^n
\; \middle| \;
\min_i \mu_i=0
\right\}
\end{equation}
\end{linenomath}
There are $n$ such faces, each having unit $\left(n-1\right)$-dimensional measure. Thus,\\
\begin{linenomath}
\begin{equation}
A_{\mathrm{admissible}}=n 
\end{equation}
\end{linenomath}
The NNC simplex satisfies
\begin{linenomath}
\begin{equation}
\Delta_{\mathrm{NNC}}
=
\left\{
\boldsymbol{\alpha}\in\left[0,1\right]^n
\; \middle| \;
\sum_{i=1}^{n}\alpha_i=1
\right\}
\end{equation}
\end{linenomath}
Its $\left(n-1\right)$-dimensional measure is
\begin{linenomath}
\begin{equation}
A_{\Delta}
=
\frac{\sqrt{n}}{\left(n-1\right)!}
\end{equation}
\end{linenomath}
By symmetry, the portion of the NNC simplex projecting to any one zero face has measure\\
\begin{linenomath}
\begin{equation}
A_{\Delta,k}
=
\frac{1}{n}
\frac{\sqrt{n}}{\left(n-1\right)!}
\end{equation}
\end{linenomath}
 
Projection along the normal direction $\mathbf{1}$ from the NNC simplex to a zero face has area-scaling factor $\sqrt{n}$. Therefore, the NNC-captured area on each zero face is\\
\begin{linenomath}
\begin{equation}
A_{\mathrm{NNC},k}
=
\sqrt{n} A_{\Delta,k}
=
\frac{1}{\left(n-1\right)!}
\end{equation}
\end{linenomath}
Since there are $n$ zero faces,
\begin{linenomath}
\begin{equation}
A_{\mathrm{NNC}}
=
\frac{n}{\left(n-1\right)!}
\end{equation}
\end{linenomath}
Hence, \\
\begin{linenomath}
\begin{equation}
\frac{A_{\mathrm{NNC}}}{A_{\mathrm{admissible}}}
=
\frac{1}{\left(n-1\right)!}
\end{equation}
\end{linenomath}
and the missed fraction is
\begin{linenomath}
\begin{equation}
\frac{A_{\mathrm{missed}}}{A_{\mathrm{admissible}}}
=
1-\frac{1}{\left(n-1\right)!}
\end{equation}
\end{linenomath}

For three objectives, NNC captures only one-half of the admissible Pareto region. More generally, the admissible capture fraction decreases factorially as $1/(n-1)!$, making the structural limitation progressively more consequential as the number of objectives increases. Thus, the geometric incompleteness of the NNC generation process is not merely a three-objective phenomenon, but an inherent structural property whose impact increases rapidly with the number of objectives. Importantly, the practical consequences of the inverse factorial relationship depend on the nature of the Pareto frontier itself.
 

\section{Formulation of the Generalized Normal Constraint Method}

The preceding sections established the invariant geometric framework induced by the multiobjective optimization problem, and demonstrated how this framework reveals the structural limitations of existing Pareto-generation methods. The purpose of this section is therefore not to introduce additional geometric constructions, but rather to formulate a Pareto-generation methodology that directly exploits the invariant geometric structures established in Section~2.

Unlike conventional algorithm development, where the methodology typically precedes geometric interpretation, the proposed Generalized Normal Constraint (GNC) method follows the opposite progression. The geometric framework is established first, independently of any particular Pareto-generation methodology. The GNC method then emerges naturally as the computational realization of these invariant geometric structures.

The proposed formulation directly adopts the complete Pareto Capture Polygon as the admissible Pareto-generation domain. A Pareto Capture Grid is then constructed over this domain, and each grid point defines the corresponding normal constraint used to generate an associated Pareto solution. The resulting methodology therefore exploits the complete Pareto Capture Polygon established by the underlying geometry, rather than a restricted subset thereof.

The formulation presented below follows directly from the invariant geometric framework developed in Section~2, and requires no additional geometric assumptions. Its implementation consists of four principal steps:


\subsection{Pareto Capture Domain}

The formulation of the GNC method begins by adopting the complete Pareto Capture Polygon established in Section~2 as the admissible Pareto-generation domain. Unlike previous deterministic methods that prescribe a particular geometric construction, GNC makes no a priori assumption regarding the shape of the admissible generation domain. Instead, the geometry is allowed to emerge directly from the invariant structures induced by the multiobjective optimization problem.

The Pareto Capture Polygon therefore constitutes the complete admissible domain over which Pareto-generation points are distributed. Because this Polygon represents the orthogonal projection of the complete Pareto Capture Polyhedron onto the Utopia Hyperplane, every admissible Pareto point possesses a unique representative point on the Pareto Capture Polygon. Consequently, constructing the generation domain over the complete Pareto Capture Polygon ensures that every admissible point is represented within the subsequent Pareto-generation process.

The GNC formulation therefore adopts the Pareto Capture Polygon without modification. This geometric construction, established independently of any particular Pareto-generation methodology, provides the complete admissible domain upon which the remainder of the GNC formulation is based.


\subsection{Pareto Capture Grid}

Once the Pareto Capture Polygon has been established as the admissible Pareto-generation domain, the next step is to construct the corresponding Pareto Capture Grid. This grid provides a discrete set of generation points from which the Pareto frontier is systematically explored.

Unlike conventional grid constructions that are prescribed independently of the underlying geometry, the Pareto Capture Grid is generated directly over the complete Pareto Capture Polygon. Consequently, the grid inherits the geometric properties of the admissible generation domain and provides an intrinsically uniform representation of the complete admissible Pareto region.

As established in Section~2, the Pareto Capture Grid is generated by projection of the zero-face representative set of the Pareto Capture Polyhedron onto the Utopia Hyperplane. This construction preserves the complete projection information while reducing the projection computation from Order $O\!\left(n_d^{\,n}\right)$ to Order $O\!\left(n_d^{\,n-1}\right)$. The resulting grid therefore provides the complete admissible generation lattice without sacrificing geometric completeness or computational efficiency.

The Pareto Capture Grid serves as the computational foundation of the GNC methodology. Each grid point represents a unique admissible projection origin on the Pareto Capture Polygon and is subsequently associated with a corresponding normal constraint for Pareto frontier generation.


\subsection{Normal Constraint Formulation}

Once the Pareto Capture Grid has been established, each Pareto Capture Grid point serves as the projection origin for the corresponding normal constraint used in Pareto frontier generation. Thus, rather than constructing normal constraints over a restricted admissible domain, as in NNC, the GNC formulation applies the same fundamental normal-constraint concept over the complete Pareto Capture Polygon.

For each Pareto Capture Grid point, the associated normal constraint is constructed using the geometric framework developed in Section~2. The corresponding optimization problem is then solved to generate the Pareto solution associated with that projection origin. Repeating this process for every point on the Pareto Capture Grid produces a representative set of Pareto solutions spanning the \textit{complete} admissible Pareto frontier.

The optimization formulation itself follows the same general structure as that employed by the NNC methodology. Consequently, the principal distinction between the two methods does not lie in the constrained optimization problem being solved, but rather in the admissible generation domain over which the normal constraints are constructed. GNC therefore preserves the computational simplicity and intuitive appeal of the normal-constraint approach while extending its geometric admissibility to the complete Pareto Capture Polygon.

The mathematical formulation associated with the $j^{th}$ Pareto Capture Grid point is therefore expressed as

\begin{linenomath}
\begin{equation}
\min_{x} \quad \bar{\mu}(x) 
\end{equation}
\end{linenomath}
subject to
\begin{linenomath}
\begin{equation}
N^{T}
\left(
\bar{\mu}(x)
-
\bar{\mu}^{\,p}_{j}
\right)
\leq
0
\end{equation}
\end{linenomath}
\begin{linenomath}
\begin{equation}
g_s(x) \leq 0  \quad (1 \leq  s \leq p)
\end{equation}
\end{linenomath}
\begin{linenomath}
\begin{equation}
h_k(x) = 0 \quad (1 \leq  k \leq q)
\end{equation}
\end{linenomath}
\begin{linenomath}
\begin{equation}
x_{li} \leq  x_i \leq x_{ui} \quad 
(1 \leq  i \leq n_{x})
\end{equation}
\end{linenomath}

\noindent where $\bar{\mu}^{p}_j$ denotes the $j^{th}$ Pareto Capture Grid point,
$\mathbf{N}$ contains the outward normal vectors defining the corresponding normal constraints,
$g$ and $h$ denote the original inequality and equality constraints of the multiobjective optimization problem.\\


\subsection{Computational Procedure}

The complete GNC methodology follows directly from the formulation presented above and consists of a sequence of computational steps that systematically generate the admissible Pareto frontier.

First, the complete Pareto Capture Polygon is adopted as the admissible Pareto-generation domain. A Pareto Capture Grid is then constructed over this domain using the zero-face representative set developed in Section~2. Each Pareto Capture Grid point defines the projection origin of the corresponding normal constraint, and the associated constrained optimization problem is subsequently solved to generate a Pareto solution.

For three-objective problems, the procedure is repeated for three representative objective orders so that each objective is minimized once. The resulting Pareto solution sets are then combined to produce a single candidate Pareto set. A Pareto filter is subsequently applied to eliminate dominated solutions. When appropriate, a proximity filter may also be employed to remove numerically redundant points while preserving the geometric representation of the Pareto frontier.

The computational procedure is summarized as follows.

\begin{enumerate}
\item Construct the Pareto Capture Polygon.
\item Generate the Pareto Capture Grid.
\item Form the corresponding normal constraints.
\item Solve the constrained optimization problem associated with each grid point.
\item Repeat for the representative objective orders.
\item Merge the resulting Pareto sets.
\item Apply Pareto filtering.
\item Apply proximity filtering, when appropriate.
\end{enumerate}

The resulting Pareto set constitutes the GNC approximation of the complete admissible Pareto frontier. The following section demonstrates the effectiveness of the proposed methodology through representative benchmark problems of increasing complexity.\\

\textit{\textbf{Two important practical points regarding the use of the word complete follow.}} While these issues are outside the scope of this work, they are briefly discussed to promote clarity.

\textit{First, in the case of \textbf{continuous Pareto frontiers}, since the Pareto set is infinite, the word complete refers to a complete representation of the Pareto set.}

\textit{Second, inevitable \textbf{numerical ill-conditioning} in certain problems may result in an incomplete numerical representation of the Pareto frontier, even in cases where complete representation is theoretically expected.}

The structural completeness established in this paper should be distinguished from numerical completeness associated with finite computations. Structural completeness is a property of the Pareto-generation methodology itself. Numerical completeness depends additionally on the numerical behavior of the optimization process. For GNC, the structural completeness established in this paper is independent of these numerical effects.

For certain problems, poor numerical conditioning may require minimizing with different objective orders, followed by merging and filtering the resulting Pareto sets, in order to achieve the theoretically expected Pareto representation—thereby enabling each method to realize its inherent structural capability: partial admissible capture for NNC and NBI, and complete admissible capture for GNC.

Historically, the Normalized Normal Constraint (NNC) method itself was introduced as an extension of the Normal Constraint (NC) method to improve numerical conditioning and the uniformity of Pareto-point distributions.

Several validation examples presented below illustrate this behavior in practice. The important point, however, is that these numerical procedures do not alter the underlying structural capability of the Pareto-generation method. Consequently, GNC remains structurally capable of generating the complete admissible Pareto frontier, whereas NNC and NBI remain structurally limited to their respective admissible subsets\\

The formulation presented in this section completes the development of the Generalized Normal Constraint methodology. As a departure from conventional developments that begin with the design of a Pareto-generation algorithm, the present methodology emerges directly from the invariant geometric framework established in Section~2. Consequently, the computational procedure preserves the geometric completeness of the admissible generation domain while retaining the simplicity and effectiveness of the normal-constraint paradigm. The following section provides computational validation of these developments.
 

\section{Computational Validation}
 
The preceding sections established the geometric framework underlying the proposed Generalized Normal Constraints (GNC) method. The purpose of the numerical studies presented here is to validate the principal theoretical claims developed throughout the paper. Accordingly, the emphasis is placed on validating the predicted geometric behavior rather than on benchmarking computational performance. Each example is not intended as an independent benchmark. Rather, each is designed to evaluate a specific aspect of the proposed framework. Collectively, the four examples progress from geometric foundations to higher-dimensional generalization, methodology, and computational applicability by addressing the following questions:

\begin{enumerate}
\item Does the proposed geometric framework achieve complete admissible Pareto capture?
\item Does the proposed geometric framework generalize naturally to higher-dimensional objective spaces?
\item Does the multi-order GNC framework behave as predicted for asymmetric Pareto frontiers?
\item Does the proposed geometric framework retain its admissible capture advantage on a fundamentally different Pareto geometry?
\end{enumerate}

Each numerical example provides direct computational validation of one of these questions, thereby establishing the correspondence between the theoretical development and the observed computational behavior of the proposed GNC framework.


\subsection{\texorpdfstring{Problem 1: Symmetric Three-Objective $p$-Norm Frontier}{Problem 1: Symmetric Three-Objective p-Norm Frontier}}

\subsubsection{Presentation and Formulations}

The first numerical example considers the three-objective $p$-norm Pareto-frontier problem
\begin{equation}
\min_{x} \; \mu_i(x)  \quad (1 \leq  i \leq n)
\end{equation}
where
\begin{equation}
\mu_i = 1 - x_i,
\qquad i=1,2,3,
\label{eq:problem1_objectives}
\end{equation}
subject to
\begin{equation}
x_1^p+x_2^p+x_3^p=1.
\label{eq:problem1_constraint}
\end{equation}
For this study, $p=4$ was selected. Two admissible design-variable bound cases were examined:
\begin{equation}
0 \leq x_i \leq 0.8,
\qquad i=1,2,3,
\label{eq:problem1_bounds_case1}
\end{equation}
and
\begin{equation}
0.2 \leq x_i \leq 0.8,
\qquad i=1,2,3.
\label{eq:problem1_bounds_case2}
\end{equation}

Because the Pareto frontier is symmetric with respect to the three objective directions, this problem isolates the effect of the proposed normal-generation geometry without introducing complications from asymmetric scaling or directional curvature. It therefore provides the baseline computational validation of the geometric framework by allowing the admissible Pareto-capture characteristics of the simplex-based NNC and projected-hypercube GNC constructions to be compared directly.

\subsubsection{NNC and GNC Pareto Frontier Results}

\begin{figure}[htbp]
\centering
\includegraphics[width=\textwidth]{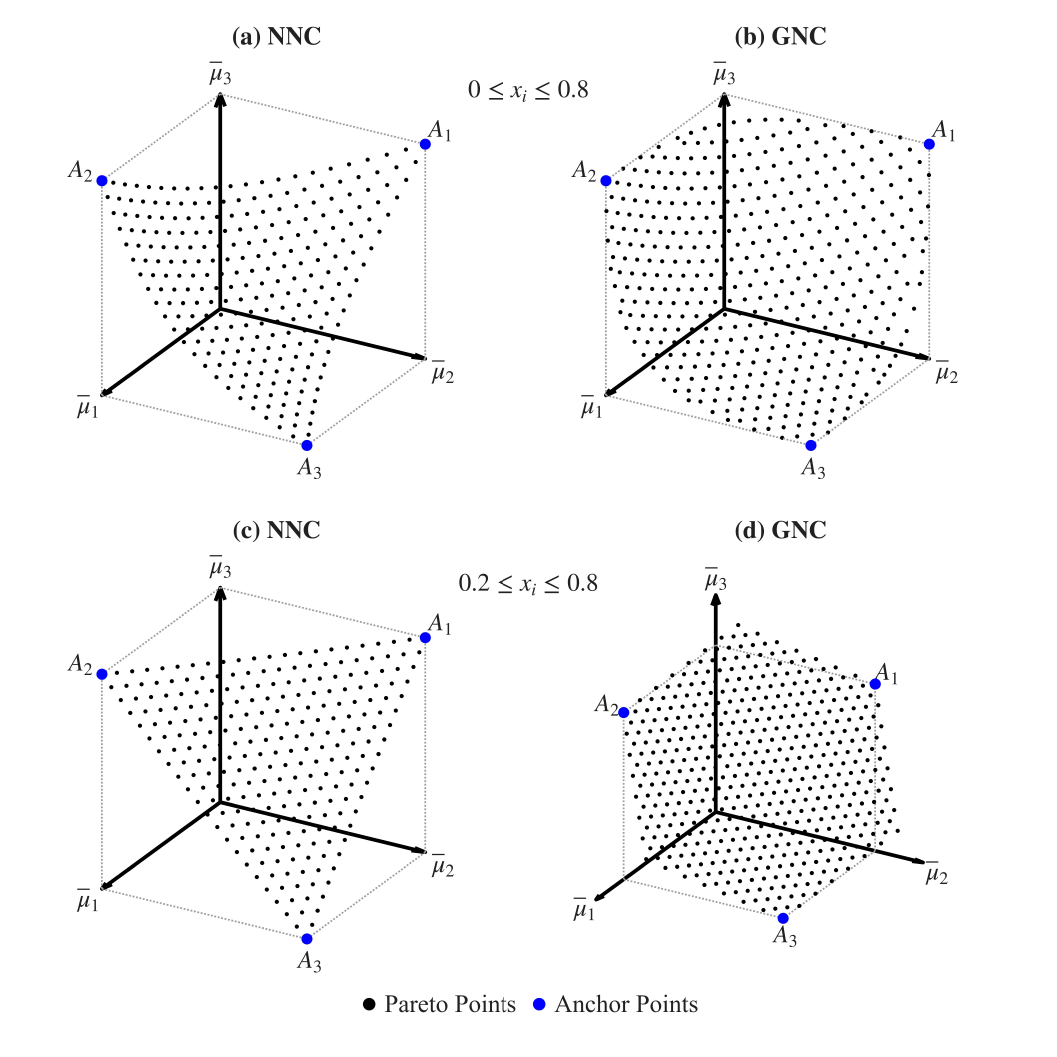}
\caption{Comparison of Pareto-frontier points generated using NNC and GNC for Problem~1. Panels (a)--(b) correspond to $0 \leq x_i \leq 0.8$, while panels (c)--(d) correspond to $0.2 \leq x_i \leq 0.8$.}
\label{fig:problem1_results}
\end{figure}

Problem~1 provides the baseline geometric validation of the GNC construction. In the first bound case, $0 \leq x_i \leq 0.8$, NNC captures the expected triangular subset associated with the simplex-based normal-generation domain. GNC extends beyond this triangular subset and captures additional admissible frontier regions surrounding the NNC triangle.

The second bound case, $0.2 \leq x_i \leq 0.8$, makes the geometric distinction more explicit. The NNC result in Fig.~\ref{fig:problem1_results}(c) captures the triangular footprint predicted by the simplex construction, while the GNC result in Fig.~\ref{fig:problem1_results}(d) captures the complete admissible hexagonal footprint. This triangle-to-hexagon transition is exactly the behavior predicted by replacing the NNC simplex with the projected-hypercube construction.

\subsubsection{Capture Characteristics}

The quantitative capture results are summarized in Table~\ref{tab:problem1_metrics}. The reported retained-point counts reflect the respective grid-generation procedures and computational implementations. They should not be interpreted as direct measures of admissible capture fraction. Rather, the admissible capture fraction is established independently through the geometric analysis and the pairwise projections.

For both methods, the number of divisions was $n_d=21$, and unique plotted points were identified using a tolerance of $10^{-6}$. The comparison is reported as the percentage increase in the number of GNC points relative to the corresponding NNC point set:

\begin{equation}
\eta_{\mathrm{increase}}
=
\frac{N_{\mathrm{GNC}}-N_{\mathrm{NNC}}}{N_{\mathrm{NNC}}}
\times 100\%.
\label{eq:problem1_gnc_point_increase}
\end{equation}

\begin{table}[htbp]
\centering
\caption{Capture-related metrics for Problem~1.}
\label{tab:problem1_metrics}
\small
\setlength{\tabcolsep}{7pt}
\begin{tabular*}{0.92\linewidth}{@{\extracolsep{\fill}}lccc}
\hline
Bound case & NNC points & GNC points & GNC increase over NNC \tabularnewline
\hline
$0 \leq x_i \leq 0.8$ & 253 & 420 & $66.0\%$ \tabularnewline
$0.2 \leq x_i \leq 0.8$ & 253 & 463 & $83.0\%$ \tabularnewline
\hline
\end{tabular*}
\end{table}

The finite-grid results in Table~\ref{tab:problem1_metrics} support the geometric observations of Fig.~\ref{fig:problem1_results}. Under the same grid setting, GNC retains $66.0\%$ more unique Pareto-frontier points than NNC in the first bound case and $83.0\%$ more in the second bound case.

These additional points are not simply a denser sampling of the same triangular region. They occupy frontier regions outside the NNC simplex-accessible subset. Thus, Problem~1 provides a direct numerical and geometric comparison of the two generation domains: NNC captures the expected triangular subset, while GNC captures the expanded admissible region associated with the projected-hypercube construction.

\subsubsection{Computational Observations}

For the complete two-case Problem~1 study, the NNC calculation required 1.500 seconds, while the GNC calculation required 5.173 seconds on the same machine and implementation. The additional computational cost is expected because GNC samples --- and produces --- a larger admissible normal-generation domain than the simplex-based NNC formulation.

The increased runtime should therefore be interpreted in the context of the expanded capture capability demonstrated in Fig.~\ref{fig:problem1_results} and Table~\ref{tab:problem1_metrics}. Rather than producing a denser sampling of the same admissible region, GNC captures additional admissible Pareto-frontier regions that are structurally inaccessible to NNC.

Therefore, Problem~1 provides the baseline computational validation of the proposed geometric framework. The projected-hypercube construction captures admissible Pareto-frontier regions that are structurally inaccessible to the simplex-based NNC construction, consistent with the theoretical development presented earlier.


\subsection{\texorpdfstring{Problem 2: Four-Objective $p$-Norm Frontier}{Problem 2: Four-Objective p-Norm Frontier}}

\subsubsection{Presentation and Formulations}

Problem~2 extends the symmetric $p$-norm frontier of Problem~1 from three objectives to four objectives. The multiobjective problem is written as
\begin{equation}
\min_{x} \; \mu_i(x)  \quad (1 \leq  i \leq n)
\end{equation}

where
\begin{equation}
\mu_i = x_i,
\qquad
i=1,2,3,4
\label{eq:problem2_objectives}
\end{equation}
subject to
\begin{equation}
\sum_{i=1}^{4}(1-x_i)^4 = 1
\label{eq:problem2_constraint}
\end{equation}
with
\begin{equation}
0 \leq x_i \leq 1,
\qquad
i=1,2,3,4
\label{eq:problem2_bounds}
\end{equation}

Since the Pareto frontier is four-dimensional, direct visualization of the complete frontier is not possible. Pairwise objective-space projections are therefore used to evaluate the retained Pareto solutions, while the analytical frontier relation in Eq.~\eqref{eq:problem2_constraint} permits direct verification of the generated points. This problem therefore provides the computational validation of the proposed geometric framework in a higher-dimensional objective space by demonstrating that the projected-hypercube construction extends naturally beyond the three-objective case.

\subsubsection{NNC and GNC Pareto Frontier Results}

\begin{figure}[htbp]
\centering
\includegraphics[width=\textwidth]{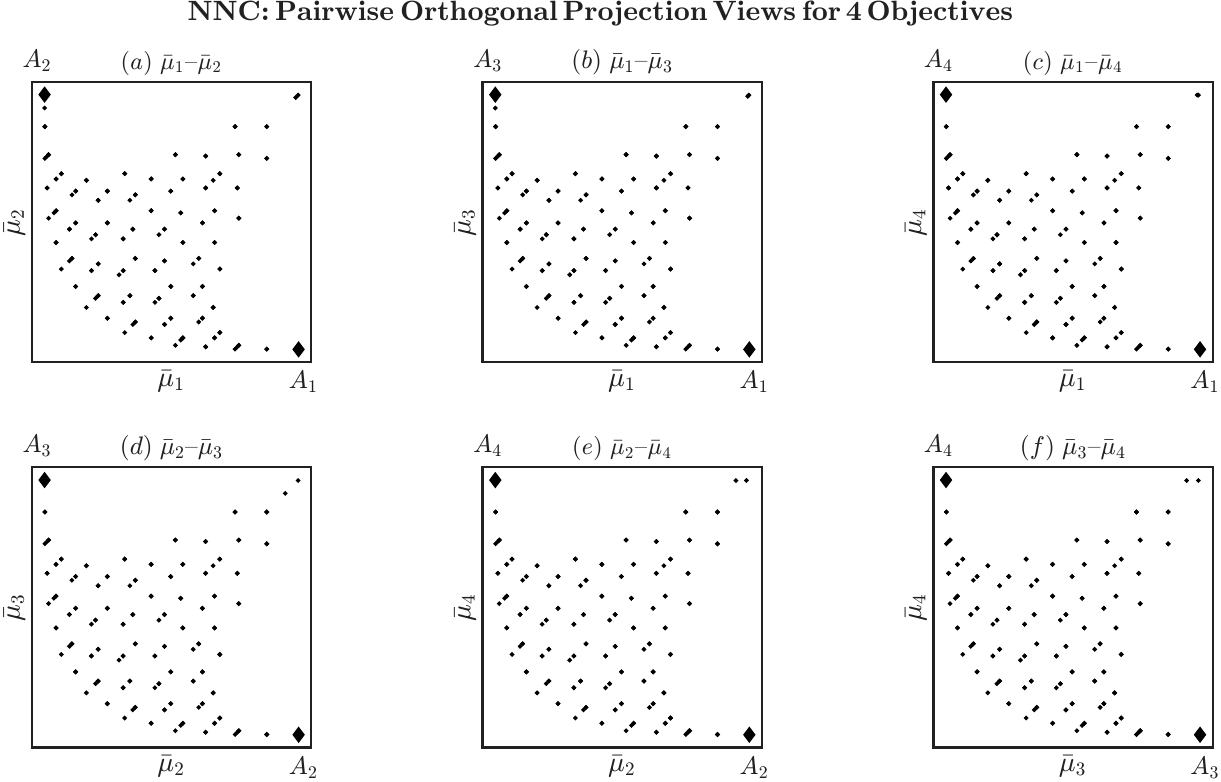}
\caption{Pairwise objective-space projections of the NNC solution for Problem~2.}
\label{fig:problem2_nnc_pairwise}
\end{figure}

\begin{figure}[htbp]
\centering
\includegraphics[width=\textwidth]{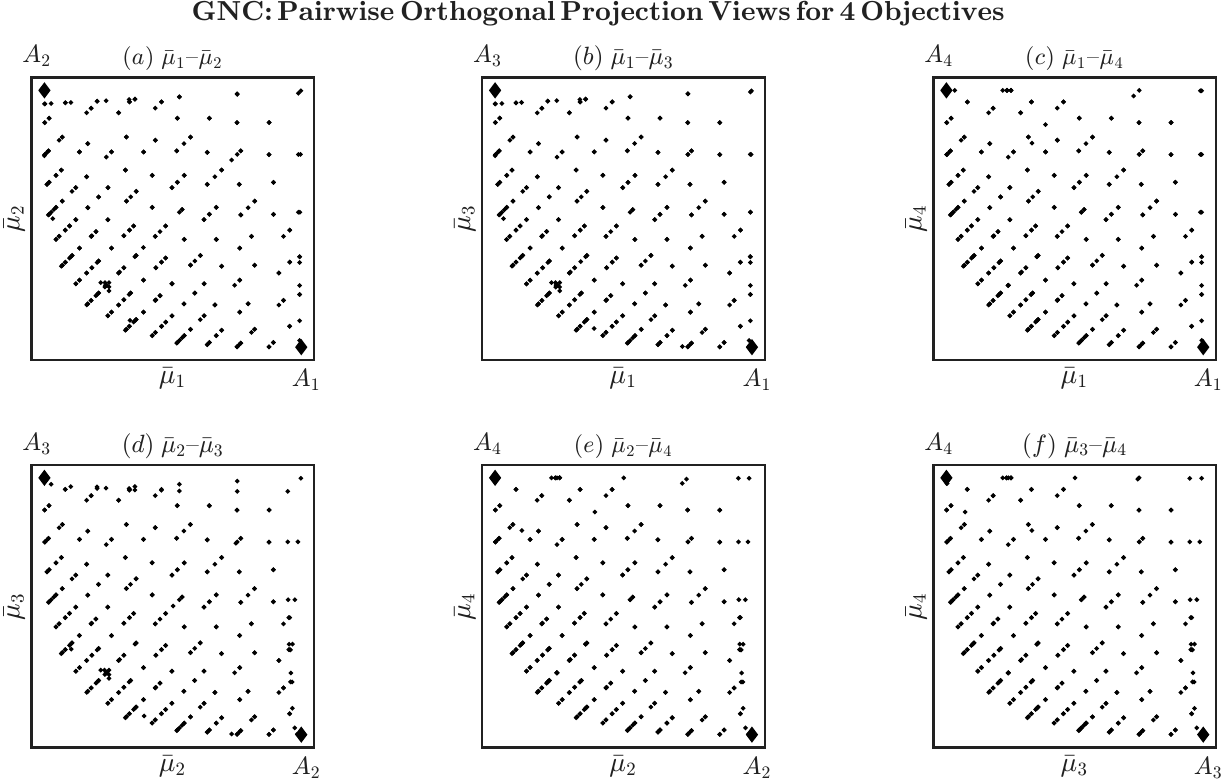}
\caption{Pairwise objective-space projections of the GNC solution for Problem~2.}
\label{fig:problem2_gnc_pairwise}
\end{figure}

Problem~2 confirms that the geometric advantage of GNC is retained in the case of four objectives. The NNC solution in Fig.~\ref{fig:problem2_nnc_pairwise} is structurally restricted and yields a subset of the admissible hyperspherical frontier. In contrast, the GNC solution in Fig.~\ref{fig:problem2_gnc_pairwise} is structurally capable of capturing the \textit{complete} admissible Pareto frontier, consistent with the pairwise projections of the subplots. The finite-grid results presented here are consistent with this theoretical property.

The important result is not simply that GNC retains more Pareto points. Rather, the additional points occupy admissible frontier regions that are inaccessible to the simplex-based NNC construction. Thus, the pairwise projections provide direct numerical evidence that GNC provides access to the complete admissible Pareto frontier for the four-objective problem.

\subsubsection{Capture and Verification Characteristics}

The principal numerical results for Problem~2 are summarized in Table~\ref{tab:problem2_metrics}. Both methods used the same grid-division setting, $n_d=8$. The reported retained-point counts reflect the respective grid-generation procedures and computational implementations. They are not direct measures of admissible capture fraction, which is established independently through the geometric analysis and the pairwise projections.

\begin{table}[htbp]
\centering
\caption{Capture and frontier-verification metrics for Problem~2.}
\label{tab:problem2_metrics}
\small
\setlength{\tabcolsep}{6pt}
\begin{tabular*}{0.92\linewidth}{@{\extracolsep{\fill}}lcc}
\hline
Metric & NNC & GNC \tabularnewline
\hline
Grid divisions, $n_d$ & 8 & 8 \tabularnewline
Raw points & 165 & 365 \tabularnewline
Pareto points & 165 & 365 \tabularnewline
Retained-point increase over NNC & -- & $121.2\%$ \tabularnewline
On-front unique points & 165 of 165 & 361 of 365 \tabularnewline
Verified-point increase over NNC & -- & $118.8\%$ \tabularnewline
Mean frontier error & $1.900\times 10^{-12}$ & $4.920\times 10^{-5}$ \tabularnewline
Maximum frontier error & $2.026\times 10^{-11}$ & $1.002\times 10^{-2}$ \tabularnewline
Mean nearest-neighbor distance & 0.1799 & 0.1612 \tabularnewline
Median nearest-neighbor distance & 0.1765 & 0.1765 \tabularnewline
Maximum nearest-neighbor distance & 0.3038 & 0.3038 \tabularnewline
\hline
\end{tabular*}
\end{table}

The finite-grid results are consistent with the theoretically expected complete admissible frontier capture observed in the pairwise projections. Under the same grid setting, GNC retains 365 Pareto points compared with 165 for NNC, corresponding to a $121.2\%$ increase in retained frontier points. When only points satisfying the analytical frontier check are counted, GNC retains 361 verified points compared with 165 for NNC, corresponding to a $118.8\%$ increase.

The frontier-verification results confirm that the expanded GNC point set remains an accurate, high-quality approximation of the analytical hyperspherical frontier while maintaining high solution quality across nearly all retained points.

The larger maximum frontier error reported for GNC is caused by a small number of boundary points generated near the outer extent of the expanded GNC domain. It does not indicate a systematic loss of accuracy across the GNC point set. This is confirmed by the mean frontier error, which remains below the prescribed tolerance.

The spacing statistics provide additional evidence of the superiority of complete admissible frontier capture rather than artificial point compression. The mean nearest-neighbor distance decreases from 0.1799 for NNC to 0.1612 for GNC, indicating a denser and more complete retained approximation. The median and maximum nearest-neighbor distances remain unchanged, showing that the added GNC points expand the represented region without creating a distorted or excessively clustered point set.

\subsubsection{Computational Observations}

The runtime increases from 0.641 seconds for NNC to 4.478 seconds for GNC. This additional computational cost is expected because the projected-hypercube construction samples a larger admissible normal-generation domain than the simplex-based NNC formulation. The additional cost is accompanied by the theoretically expected complete admissible frontier representation demonstrated in Figs.~\ref{fig:problem2_nnc_pairwise}--\ref{fig:problem2_gnc_pairwise} and supported by the metrics in Table~\ref{tab:problem2_metrics}.

Therefore, Problem~2 confirms that the proposed geometric framework generalizes naturally to higher-dimensional objective spaces. The GNC generation domain captures the complete and verified approximation of the admissible hyperspherical Pareto frontier, while NNC remains limited to the simplex-accessible subset.


\subsection{\texorpdfstring{Problem 3: Skewed Three-Objective $p$-Norm Frontier With Numerical Order Dependence}{Problem 3: Skewed Three-Objective p-Norm Frontier}}

\subsubsection{Presentation and Formulations}

The third numerical example considers a skewed three-objective $p$-norm Pareto-frontier problem. The multiobjective problem is written as
\begin{equation}
\min_{x} \; \mu_i(x)  \quad (1 \leq  i \leq n)
\end{equation}
\begin{equation}
\mu_1 = 1-x_1,
\qquad
\mu_2 = 1-x_2,
\qquad
\mu_3 = 1-x_3
\label{eq:problem3_objectives}
\end{equation}
subject to
\begin{equation}
x_1^2+x_2^4+x_3^8=1
\label{nonLConProb3}
\end{equation}
with
\begin{equation}
0 \leq x_i \leq 1,
\qquad
i=1,2,3
\label{eq:problem3_bounds}
\end{equation}

Unlike the symmetric frontier of Problem~1, this formulation introduces unequal curvature in the three objective directions through the exponents $[2,4,8]$. The Pareto frontier is therefore no longer geometrically balanced about the anchor points. As a result, different objective orders naturally produce complementary approximations of the same admissible Pareto frontier. This problem therefore provides the computational validation of the proposed multi-order GNC framework by demonstrating how these complementary approximations can be consolidated into a complete frontier representation through merging and proximity filtering.

\subsubsection{Objective-Order Dependence}

\begin{figure}[htbp]
\centering
\includegraphics[width=\textwidth]{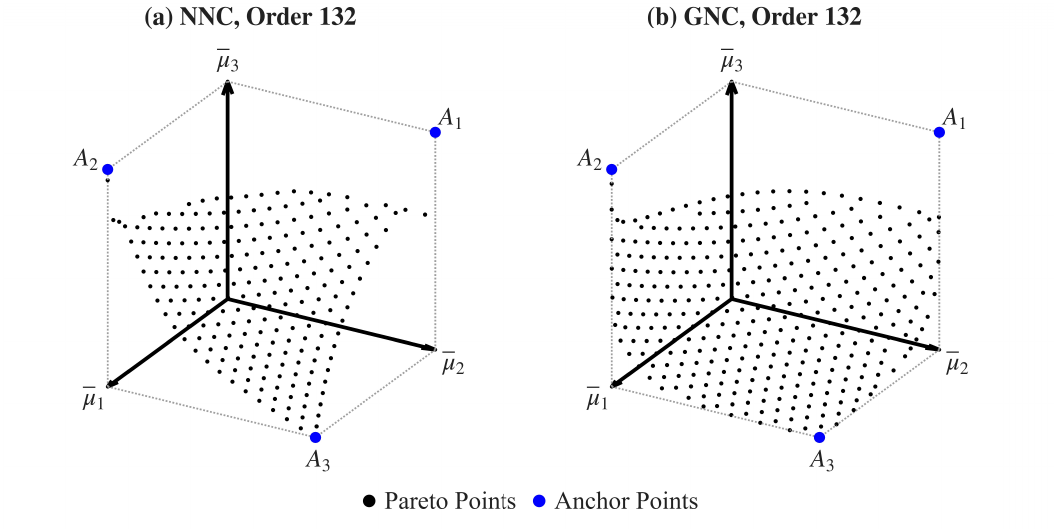}
\caption{Pareto-frontier points generated using NNC and GNC for Problem~3 with objective order 132.}
\label{fig:problem3_order132}
\end{figure}

\begin{figure}[htbp]
\centering
\includegraphics[width=\textwidth]{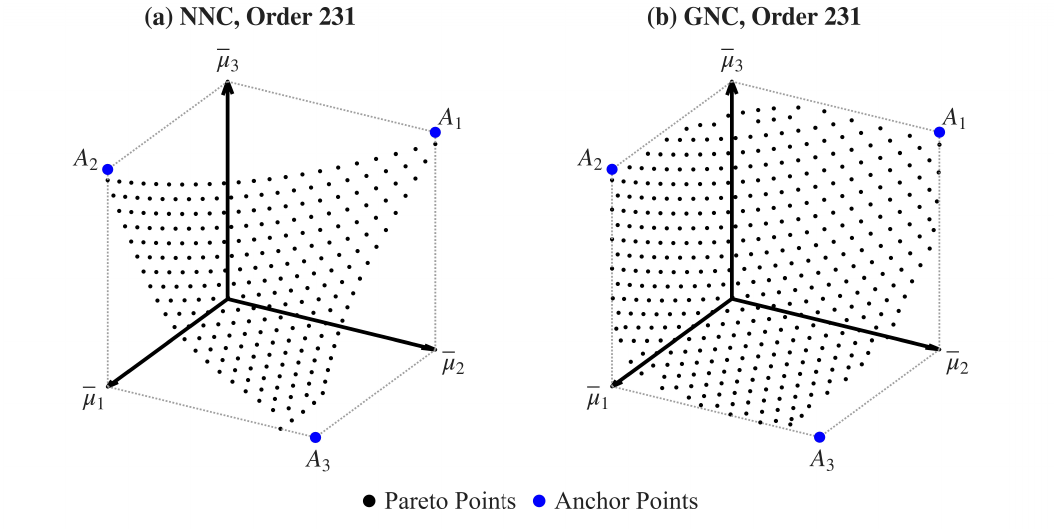}
\caption{Pareto-frontier points generated using NNC and GNC for Problem~3 with objective order 231.}
\label{fig:problem3_order231}
\end{figure}

Problem~3 confirms that skewed frontier geometry can produce order-dependent frontier approximations. Because the constraint in Eq.~\ref{nonLConProb3} contains the unequal exponents $2$, $4$, and $8$, the frontier curvature varies significantly from one objective direction to another. As a result, different objective orders produce complementary approximations of the asymmetric Pareto frontier.

This behavior is evident in Figs.~\ref{fig:problem3_order132} and~\ref{fig:problem3_order231}. Different objective orders generate distinct, but complementary, approximations of the asymmetric frontier. The observed differences are not numerical anomalies; they arise from the interaction between the objective-order construction and the asymmetric frontier geometry.

For both objective orders, GNC produces the broader approximation. The NNC points remain confined to a more limited subset of the admissible frontier, whereas the GNC points extend farther toward the boundary regions and capture larger admissible portions of the tradeoff surface. Thus, the order-dependent results show that the projected-hypercube construction remains advantageous even when the frontier geometry is no longer symmetric.

\subsubsection{Merged Frontier Representation and Proximity Filtering}

\begin{figure}[htbp]
\centering
\includegraphics[width=\textwidth]{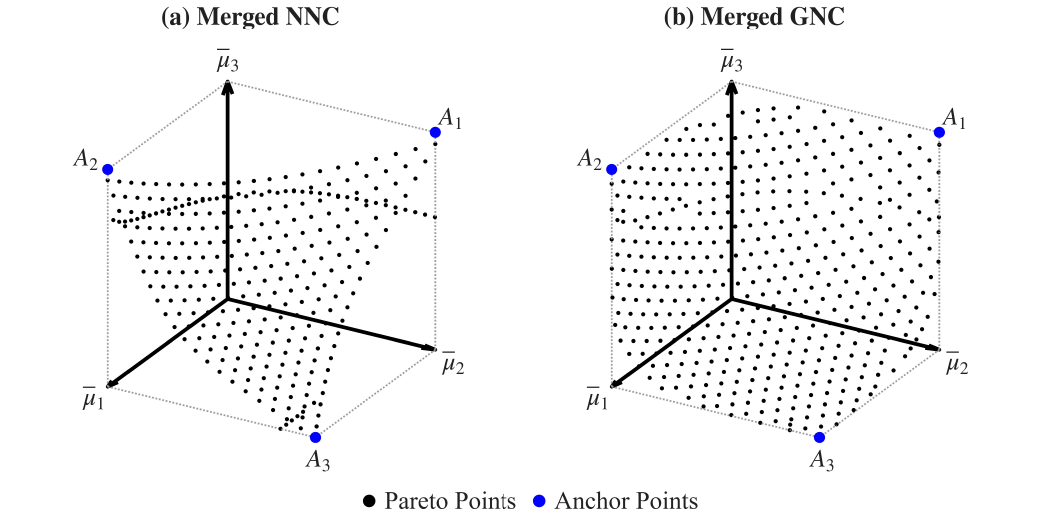}
\caption{Pareto-frontier points generated using NNC and GNC for Problem~3 after merging objective orders 132 and 231.}
\label{fig:problem3_merged}
\end{figure}

The merged result in Fig.~\ref{fig:problem3_merged} consolidates complementary frontier information from multiple valid objective-order runs. For this skewed problem, merging is a natural requirement of the geometry: different orderings emphasize different admissible regions of the same Pareto frontier. Merging the objective-order runs produces the theoretically expected representation of the admissible frontier without altering the underlying geometry.

A noticeable feature of the merged point sets is the accumulation of points near some frontier edges. These points are not numerical artifacts. They arise because different objective-order runs can generate identical or extremely close solutions in relatively flat regions of the frontier. To prevent these redundant edge clusters from overstating the number of distinct retained solutions, duplicate and near-duplicate points are removed using the stated proximity tolerance. This filtering step preserves the non-dominated frontier structure while producing a cleaner comparison of distinct frontier representations.

The benefit of this process is clear in Fig.~\ref{fig:problem3_merged}. After complementary objective-order-dependent runs are merged and redundant points are removed, the final GNC result retains the theoretically expected complete representation of the skewed frontier. The corresponding NNC result remains more restricted even after the same merging and filtering procedure, illustrating that merging addresses objective-order dependence while preserving the underlying geometric distinction between the two methods.

\subsubsection{Capture Characteristics}

The quantitative results for Problem~3 are summarized in Table~\ref{tab:problem3_metrics}. All cases used $n_d=21$ for both methods, and unique plotted points were identified using a tolerance of $10^{-6}$. The final column reports the percentage increase in unique GNC points relative to NNC.

\begin{table}[htbp]
\centering
\caption{Capture-related metrics for Problem~3 after duplicate and near-duplicate points are removed using the stated unique-point tolerance.}
\label{tab:problem3_metrics}
\small
\setlength{\tabcolsep}{7pt}
\begin{tabular*}{0.92\linewidth}{@{\extracolsep{\fill}}lccc}
\hline
Case & NNC points & GNC points & GNC increase over NNC \tabularnewline
\hline
Order 132 ($\mu_2$ minimized) & 218 & 336 & $54.1\%$ \tabularnewline
Order 231 ($\mu_1$ minimized) & 249 & 406 & $63.1\%$ \tabularnewline
Merged orders 132 and 231 & 259 & 438 & $69.1\%$ \tabularnewline
\hline
\end{tabular*}
\end{table}

The results confirm two important features of the skewed problem. First, the frontier representation depends on the selected objective order, as expected for an asymmetric frontier. Second, for every case considered, GNC retains a substantially larger set of distinct Pareto-frontier points than NNC. The GNC increase ranges from $54.1\%$ for objective order 132 to $69.1\%$ for the merged case.

The merged case is especially instructive. Before filtering, the merged NNC result contained 505 retained points, but these reduced to 259 unique plotted points after duplicate and near-duplicate solutions were removed. By contrast, the merged GNC result retained 438 unique plotted points under the same tolerance. This result confirms that the complete GNC representation is not an artifact of redundant sampling; it reflects a complete admissible frontier representation.

\subsubsection{Computational Observations}

The computational results for Problem~3 are summarized in Table~\ref{tab:problem3_runtime}. The table reports both runtime and the number of unique plotted points, since the main computational question is whether the additional cost of GNC and merging is accompanied by a meaningful increase in distinct retained frontier points.

\begin{table}[htbp]
\centering
\caption{Runtime and unique-point metrics for Problem~3.}
\label{tab:problem3_runtime}
\small
\setlength{\tabcolsep}{8pt}
\begin{tabular*}{0.90\linewidth}{@{\extracolsep{\fill}}lcccc}
\hline
Case & $t_{\mathrm{NNC}}$ (s) & $t_{\mathrm{GNC}}$ (s) & $N_{\mathrm{NNC}}$ unique & $N_{\mathrm{GNC}}$ unique \tabularnewline
\hline
Order 132 & 0.703 & 2.352 & 218 & 336 \tabularnewline
Order 231 & 0.615 & 2.406 & 249 & 406 \tabularnewline
Merged 132 and 231 & 1.308 & 4.892 & 259 & 438 \tabularnewline
\hline
\end{tabular*}
\end{table}

As expected, GNC requires more computational time than NNC for each objective-order case because it samples and produces the complete projected-hypercube generation domain. The additional cost is accompanied by consistently larger sets of distinct retained Pareto-frontier points: GNC produces $54.1\%$ more unique points for order 132, $63.1\%$ more for order 231, and $69.1\%$ more for the merged case.

The merged results also show why proximity filtering is important for skewed frontiers. Merging multiple objective-order runs increases the amount of frontier information available, but it can also introduce duplicate or nearly duplicate points, particularly near frontier edges where different orderings recover similar solutions. Proximity filtering removes these redundant points without changing the underlying Pareto-frontier structure.

Therefore, Problem~3 demonstrates that asymmetric frontier geometry produces complementary objective-order-dependent approximations of the admissible Pareto frontier. Merging and proximity filtering provide a practical means of realizing the theoretically expected frontier representation without altering the underlying structural capability of the Pareto-generation method. Thus, NNC realizes its theoretically expected partial admissible capture, whereas GNC realizes its theoretically expected complete admissible capture.

\subsection{Problem 4: Spherical Frontier Benchmark}

\subsubsection{Presentation and Formulations}

The fourth numerical example considers a spherical Pareto-frontier benchmark. The objective functions are defined as
\begin{equation}
\min_{x} \; \mu_i(x)  \quad (1 \leq  i \leq n)
\end{equation}
\begin{equation}
\mu_1 = x_1,
\qquad
\mu_2 = x_2,
\qquad
\mu_3 = x_3
\label{eq:problem4_objectives}
\end{equation}
subject to
\begin{equation}
x_1^2+x_2^2+x_3^2 \leq 1
\label{eq:problem4_constraint}
\end{equation}
with
\begin{equation}
x_1 \geq -1,
\qquad
x_2 \geq -1,
\qquad
x_3 \geq -0.5
\label{eq:problem4_bounds}
\end{equation}

Unlike the preceding $p$-norm examples, the objective functions coincide directly with the design variables. The geometry of the Pareto frontier is therefore especially transparent. This problem provides computational validation of the proposed geometric framework for a fundamentally different Pareto geometry by demonstrating that the projected-hypercube construction retains its admissible capture advantage beyond the $p$-norm benchmark family.

\subsubsection{NNC and GNC Pareto Frontier Results}

\begin{figure}[htbp]
\centering
\includegraphics[width=\textwidth]{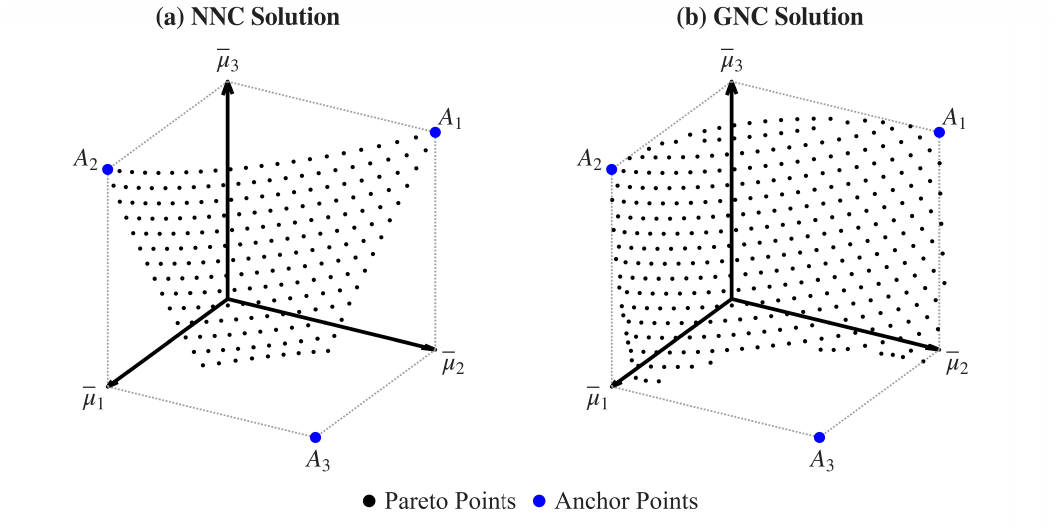}
\caption{Comparison of Pareto-frontier points generated using NNC and GNC for Problem~4.}
\label{fig:problem4_results}
\end{figure}

Problem~4 confirms that the geometric capture advantage of GNC is not limited to the preceding $p$-norm examples. Both methods generate Pareto-frontier points on the spherical surface and identify the same anchor-point structure. The difference lies in the portion of the admissible spherical surface reached by each normal-generation domain.

The NNC solution in Fig.~\ref{fig:problem4_results} forms a well-defined but restricted triangular approximation of the spherical frontier. In contrast, the GNC solution extends beyond this triangular patch and captures the complete admissible portion of the same Pareto surface. The additional GNC points follow the curvature of the spherical frontier toward the boundary regions, consistent with the expanded projected-hypercube construction.

\subsubsection{Capture Characteristics}

Problem~4 reinforces the distinction between point density and admissible frontier capture. Increasing the number of NNC sampling locations would make the triangular patch denser, but it would not change the underlying portion of the spherical frontier that is accessible to the simplex-based construction. The limitation is geometric, not merely numerical.

The quantitative results are summarized in Table~\ref{tab:problem4_metrics}. Both methods used $n_d=21$. GNC retained 346 Pareto-frontier points compared with 217 for NNC, corresponding to a $59.4\%$ increase in retained points under the same grid setting. The reported retained-point counts reflect the respective grid-generation procedures and computational implementations. They are not direct measures of admissible capture fraction. Rather, the admissible capture fraction is established independently through the geometric analysis and graphical comparison.

\begin{table}[htbp]
\centering
\caption{Capture-related and runtime metrics for Problem~4.}
\label{tab:problem4_metrics}
\small
\setlength{\tabcolsep}{8pt}
\begin{tabular*}{0.82\linewidth}{@{\extracolsep{\fill}}lcc}
\hline
Metric & NNC & GNC \tabularnewline
\hline
Grid divisions, $n_d$ & 21 & 21 \tabularnewline
Retained points & 217 & 346 \tabularnewline
Retained-point increase over NNC & -- & $59.4\%$ \tabularnewline
Runtime (s) & 0.840 & 3.890 \tabularnewline
\hline
\end{tabular*}
\end{table}

These additional points are not produced by a different Pareto definition or a different problem formulation. They arise because the GNC generation domain includes admissible normal directions outside the NNC simplex. Thus, the figure and the accompanying metrics provide complementary geometric and computational evidence of the admissible frontier-capture distinction between NNC and GNC.

\subsubsection{Computational Observations}

The runtime increases from 0.840 seconds for NNC to 3.890 seconds for GNC. This additional computational cost is expected because the projected-hypercube construction samples a larger admissible normal-generation domain than the simplex-based NNC formulation. The additional computational cost is accompanied by the theoretically expected complete admissible frontier representation demonstrated in Fig.~\ref{fig:problem4_results} and supported by the metrics in Table~\ref{tab:problem4_metrics}.

Therefore, Problem~4 confirms that the proposed geometric framework is not limited to $p$-norm frontiers. The projected-hypercube construction continues to realize the theoretically expected complete admissible Pareto representation even for a fundamentally different spherical geometry, while the simplex-based NNC formulation remains restricted to a smaller admissible subset.


\subsection{Unifying Observations}

The four numerical examples provide computational validation of the principal theoretical developments presented in this paper. Problem~1 validates the baseline admissible-capture geometry of the proposed framework. Problem~2 demonstrates that the projected-hypercube construction generalizes naturally to higher-dimensional objective spaces. Problem~3 validates the multi-order GNC framework for asymmetric Pareto frontiers by showing that complementary objective-order-dependent approximations can be consolidated through merging and proximity filtering. Finally, Problem~4 confirms that the geometric framework is not limited to the $p$-norm benchmark family, but remains effective for a fundamentally different spherical Pareto geometry.

A consistent geometric interpretation emerges across all four examples. NNC produces structured and repeatable Pareto-frontier approximations, but those approximations remain confined to the simplex-accessible subset of the admissible frontier. Increasing the NNC grid density of retained points within that subset does not alter the underlying normal-generation domain. In contrast, GNC replaces the simplex with the projected hypercube, thereby expanding the admissible normal-generation domain to the complete Pareto Capture Polygon and enabling the complete admissible Pareto representation.

Although the specific manifestations differ among the benchmark problems, the underlying geometric mechanism remains unchanged. In the symmetric three-objective problem, it appears as the transition from the NNC triangle to the GNC hexagon. In the four-objective problem, the same mechanism extends naturally to higher-dimensional objective space and is supported by analytical frontier verification. In the asymmetric benchmark, Pareto frontier flatness introduces objective-order dependence, making complementary orderings beneficial; merging and proximity filtering then consolidate these complementary approximations into the theoretically expected complete frontier representation. Finally, the spherical benchmark demonstrates that the same geometric behavior is maintained for a fundamentally different Pareto geometry.

The computational results are consistent with this interpretation. GNC requires additional computational effort because it samples a larger admissible normal-generation domain. However, the additional cost is consistently accompanied by a substantial increase in distinct retained Pareto-frontier points and complete frontier representation, rather than a denser sampling of the same simplex-accessible region.

Overall, the numerical studies validate the central geometric contribution of this paper. The proposed projected-hypercube framework removes the geometric limitations imposed by the simplex-based NNC construction, thereby enabling the theoretically expected complete representation of the admissible Pareto frontier across the benchmark problems considered.


\section{Concluding Remarks}
This paper developed the Generalized Normal Constraint (GNC) framework as a unified geometric and computational methodology for complete admissible Pareto frontier generation. Rather than beginning with the formulation of a new optimization algorithm, the proposed approach first identified \textit{the invariant geometric structures induced by the multiobjective optimization problem itself}, from which the GNC methodology naturally emerged. The resulting framework provides complete admissible Pareto capture while preserving the simplicity and intuitive appeal of the normal-constraint paradigm. Theoretical developments established the geometric completeness of the proposed framework and quantified the structural incompleteness of NNC through a closed-form inverse-factorial relationship. Computational studies further confirmed the theoretical predictions across representative benchmark problems. Collectively, these developments establish GNC as a comprehensive geometric generalization of the NNC methodology and provide a unified foundation for future developments in deterministic Pareto frontier generation.

\section*{Author Contributions}

 Messac conceived the GNC framework, developed the theoretical formulation, performed the mathematical derivations, supervised the research, and wrote the manuscript. Montaque coauthored the Introduction, Results, and References Sections; and contributed to software implementation, computational studies, validation, figure development, and manuscript review. Both authors have read and agreed to the published version of the manuscript.

\section*{Funding}
This research used no external funding.

\section*{Ethical Approval}
Not applicable.

\section*{Informed Consent}
Not applicable.

\section*{Data Availability}
The study is primarily methodological and did not involve external datasets. Custom MATLAB codes were developed for the computational examples.

\section*{Acknowledgments}
The authors acknowledge Howard University for supporting this research effort.

\section*{Conflict of Interest}
The authors declare no conflicts of interest.

\section*{Abbreviations}
The following abbreviations are used in this manuscript:

\noindent
\begin{tabular}{@{}ll}
GNC & Generalized Normal Constraint\\
NC & Normal Constraint\\
NNC & Normalized Normal Constraint\\
NBI & Normal-Boundary Intersection\\
ENNC & Enhanced Normalized Normal Constraint\\
DSD & Directed Search Domain\\
MOOP & Multiobjective Optimization Problem\\
SOOP & Single-Objective Optimization Problem\\
\end{tabular}

 
\bibliography{references}

\clearpage
 
\end{document}